\documentclass[preprint, amsmath, amssymb, amsfonts]{revtex4-1}
\pdfoutput=1
\usepackage{graphicx}
\usepackage[font={small,bf}]{caption}
\usepackage{comment}
\usepackage{amssymb}
\usepackage{amsmath}
\usepackage{subfig}
\usepackage{enumitem}
\newlist{steps}{enumerate}{1}
\setlist[steps, 1]{label = Step \arabic*:}
\begin{document}

\title{Adaptive Pressure Control for Use in Variable-Thrust Rocket Development}

\author{Anil Alan}
\affiliation{Bilkent University, Ankara, Turkey}
\author{Yildiray Yildiz}
\affiliation{Bilkent University, Ankara, Turkey}
\author{Umit Poyraz}
\affiliation{Roketsan INC., Ankara, Turkey}

\begin{abstract}
The precise control of gas generator pressure is important to obtain variable thrust in ducted rockets. A delay resistant closed loop reference model adaptive control is proposed in this paper to address this problem. The proposed controller combines delay compensation and adaptation with improved transient response. The controller is successfully implemented using an industrial grade cold air test setup, which is a milestone towards obtaining a fully developed throttleable rocket gas generator controller. Simulation and experimental comparisons with alternative adaptive approaches and a fixed controller demonstrate improved performance and effective handling of time delays and uncertainties. A step by step design methodology, covering robustfying schemes, selection of adaptation rates and initial controller parameters, is also provided to facilitate implementations. 
\end{abstract}

\maketitle

\section{Introduction}

\begin{figure} [htp]
	\centering
	\includegraphics[trim = 30mm 70mm 40mm 50mm, clip, width=15cm]{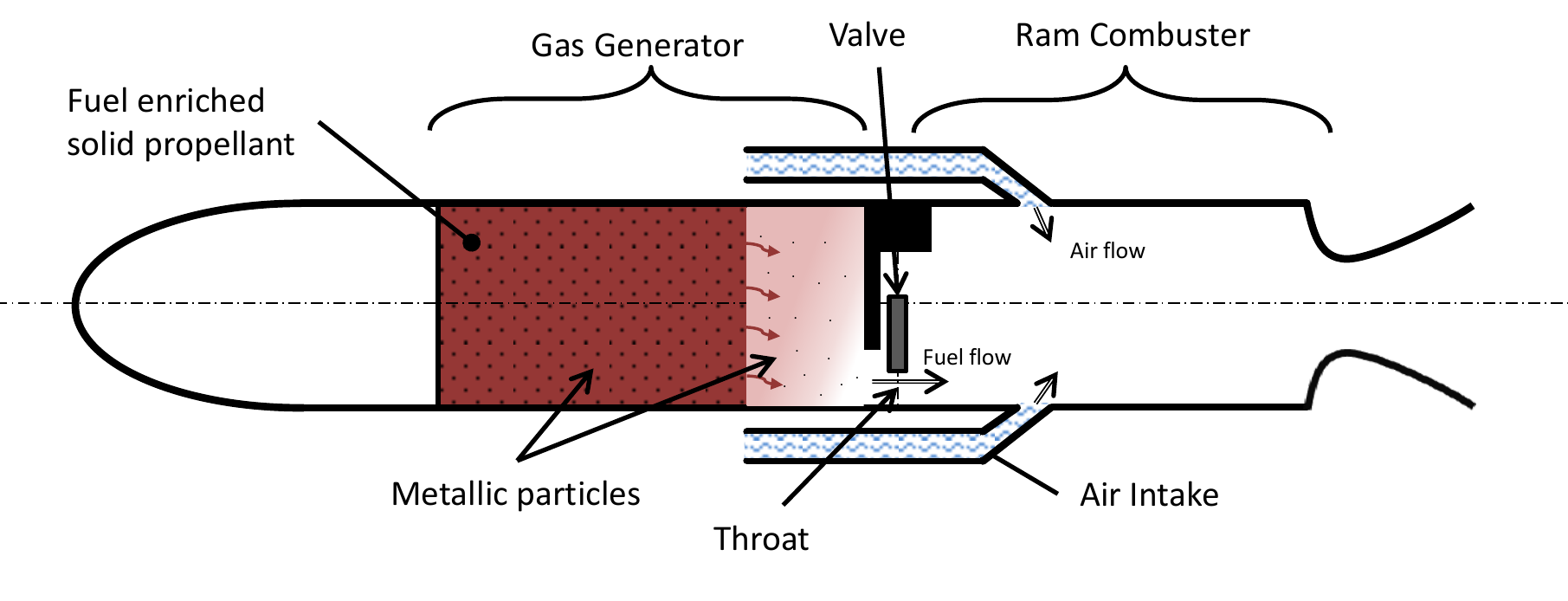}
	\caption{Throttleable ducted rocket components}
	\label{fig:TDR}
\end{figure}

Precise control of the gas pressure in the gas generator is the key element of having variable thrust for a ducted rocket, and a great example of a general control problem where time delays, uncertainties and nonlinearities need to be carefully addressed. A ducted rocket whose thrust is controlled is called a throttleable ducted rocket (TDR, see Fig. \ref{fig:TDR}). Controlling the thrust gives the ability to alter the speed during a mission and allow the missile to make maneuvers more efficiently. This way, the missile enlarges its no-escape zone, which is the maximum range that it can outrun its target \cite{Besser:12}. 

As seen in Figure \ref{fig:TDR}, TDR propulsion systems contain two main parts, the gas generator (GG) and the ram combuster (RC). Fuel-rich (or oxidizer deficient) solid propellant in GG is ignited and partially burned to obtain fuel in the gaseous form which is then sent to the RC to meet with air taken in and compressed by the air-intakes and have further combustion to produce thrust. Variable thrust is achieved by modifying the fuel mass flow rate from the GG to the RC, which can be done using several different methods including changing the burning area of the GG propellant in a controlled manner, changing the throat area between the GG and the RC using a mechanical valve (see Fig. \ref{fig:TDR}), introducing secondary injection to the GG chamber to control burning rate and utilizing a vortex valve which introduces a swirl to the flow in order to control the effective throat area \cite{Goldman,Miller:81}. Among these methods of regulating the fuel mass flow rate, changing the throat area using a mechanical valve is the most commonly used approach \cite{Davis:03,Chang:11,Burroughs,Ostrander:97,Miller:81,Bao:10,Bao:11,Chang:14}.

\begin{figure} [htp]
	\centering
	\includegraphics[trim = 0mm 60mm 10mm 10mm, clip, width=17cm]{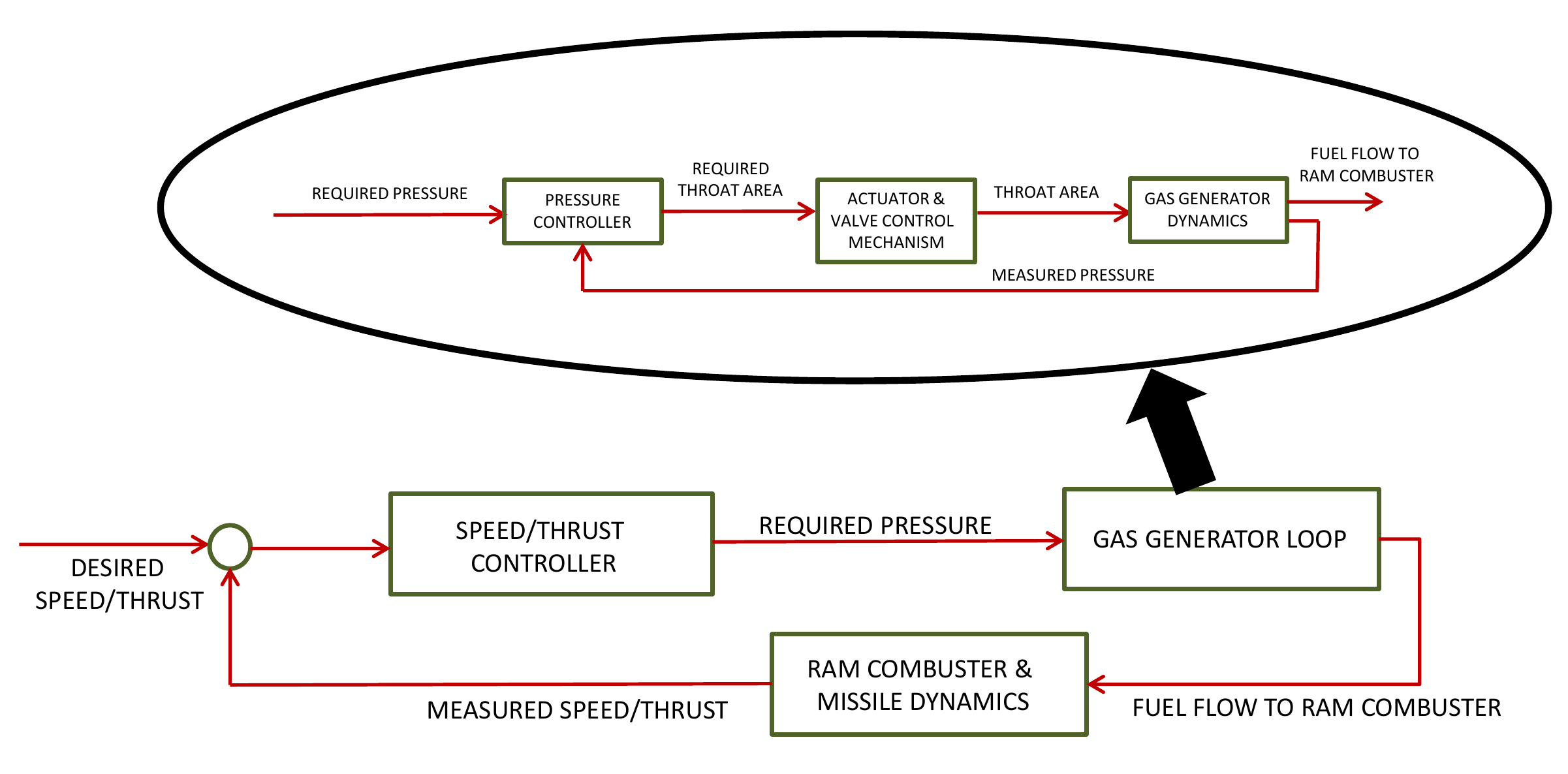}
	\caption{Thrust/speed control loop hierarchical structure}
	\label{fig:thrustloop}
\end{figure}

A typical thrust/speed control loop has a hierarchical structure where an outer loop is driven by the error between the desired and the measured thrust/speed and determines the required GG pressure to minimize this error. Required GG pressure is then provided to the inner feedback loop as a reference, which controls the mass flow rate to achieve the desired pressure (see Fig. \ref{fig:thrustloop}) \cite{Besser:12,Thomaier:87,Sreeriatha:99,Pinto:11}. The GG pressure control loop is the focus of this paper. Studies in the literature on the GG pressure control problem mainly involves linearization of the nonlinear system dynamics around certain operating points and designing linear controllers for the resultant linear dynamics.  Among others, proportional integral (PI) type controllers appear to be the most common controller used for this purpose \cite{Sreeriatha:99,Bao:2010,Niu:10}. Another challenge for the GG control problem, apart from the nonlinear dynamics, is that as the fuel burns, the free volume inside GG increases with time, which makes the system time varying. Moreover, the GG dynamics contain several uncertainties emanating from metallic particles inside the solid fuel, such as deposition at nozzle throat and ablation of mechanical elements. One of the more sophisticated control approaches to address these issues is gain scheduling \cite{Pinto:11,Thomaier:87}. In this approach, a full knowledge of the controlled plant in terms of a high fidelity mathematical model is employed to prepare look-up tables which are then used to assign appropriate controller gains at different operation modes and conditions. Another method, which is employed for the flight performance evaluation study of the Meteor missile, is called as the `performance funnel' \cite{Bauer:12,Ilchmann:02}. In this approach, a proportional controller is utilized with a time varying gain, which is adjusted online to keep the error of the closed loop system within a predefined performance funnel. 

In this study, we build upon the above mentioned successful approaches by eliminating the need for a precise system model for classical approaches and eliminating the conservatism introduced by robust methods. This is achieved by developing and utilizing a delay resistant adaptive controller which includes a unique combination of the elements of the Adaptive Posicast Controller (APC) studied by the author Yildiz and his coworkers in \cite{Yil:Auto} and closed loop reference model (CRM) adaptive controller proposed in \cite{Gibson:12,Gibson:12_1,Gibson:12_3}.

APC is an adaptive controller developed for time delay systems which extends the ideas from the Smith Predictor \cite{Smith}, finite spectrum assignment \cite{Manitius} and adaptation \cite{Ichikawa,Ortega,Sil:03}. The ability of the APC to accommodate large delays has been successfully validated through several simulation and experimental studies presented in \cite{YilJ:idle,YilJ:FAR,Yil:07,Yil:08,Yil:ACC08,Dydek:13}. Other notable studies on the adaptive control of time delay systems can be seen in \cite{Pietri:auto}, where unknown input delays and \cite{Liberis}, where both state and input delays are addressed. Also, extension of predictor feedback to nonlinear and delay adaptive systems with actuator dynamics modeled by partial differential equations can be found in \cite{Krstic:book}. Closed loop reference model (CRM) adaptive control proposed in \cite{Gibson:12,Gibson:12_1,Gibson:12_2,Gibson:12_3,Gibson:12_4} introduces an error feedback modification of the reference model which is shown to improve the transient response. Similar approaches where the reference model is modified to obtain better transients can be found in \cite{Lavretsky:12,Stepanyan:10,Stepanyan:11,Yucelen:14}. Therefore, in this paper, APC and CRM adaptive control constitute the main ingredients of the proposed controller design which can handle large delays and provide improved transient response. Furthermore, the projection algorithm \cite{Pomet:92} is used to prevent parameter drift and a procedure to select the adaptation and CRM feedback gains, inspired from \cite{YilJ:idle} and \cite{Gibson:12}, are provided to reduce the tuning time and effort for experimental implementations. The effectiveness of this approach is validated through extensive simulations together with experiments conducted using an industry grade cold air test setup, where comparisons with conventional MRAC and CRM adaptive control are provided. 

In TDR research, cold air test setup (CATS) is widely appointed as the crucial step in validation of subsystems and methods. CATS is used, for example, to validate the numerical simulation results of flow characteristics, to test the structures that are used to change the throat area and to characterize the materials that are planned to be used in the construction \cite{Lee:13catp,Deng:15,Heo:15,Ko:13,Lee:08, Verma:11}. Controller is also an important subsystem that is required to be qualified and CATS is utilized to conduct comparative analysis of alternative control systems \cite{Peterson:12}, which is then used to acquire a proper control methodology based on the gained insight. In this study, using the facilities provided by Roketsan Inc., we setup and utilize an industrial grade CATS to validate the proposed controller and compare it by various competing methods. 

Organization of the paper is as follows: Section \ref{sec:model} describes the mathematical modeling of the system together with model enhancement steps using experimental data. In Section \ref{sec:controller}, the designs of MRAC, CRM and the proposed DR-CRM adaptive controllers and a PI controller are described with implementation enhancements. Simulation results are presented in Section \ref{sec:sim} for two different scenarios. Experimental setup is described and experiment results for the same scenarios are given in Section \ref{sec:experiment}. A summary is provided in Section \ref{sec:conc}.

\section{System Model} \label{sec:model}
\begin{figure} [htp]
	\centering
	\includegraphics[trim = 30mm 80mm 60mm 20mm, clip, width=1\textwidth]{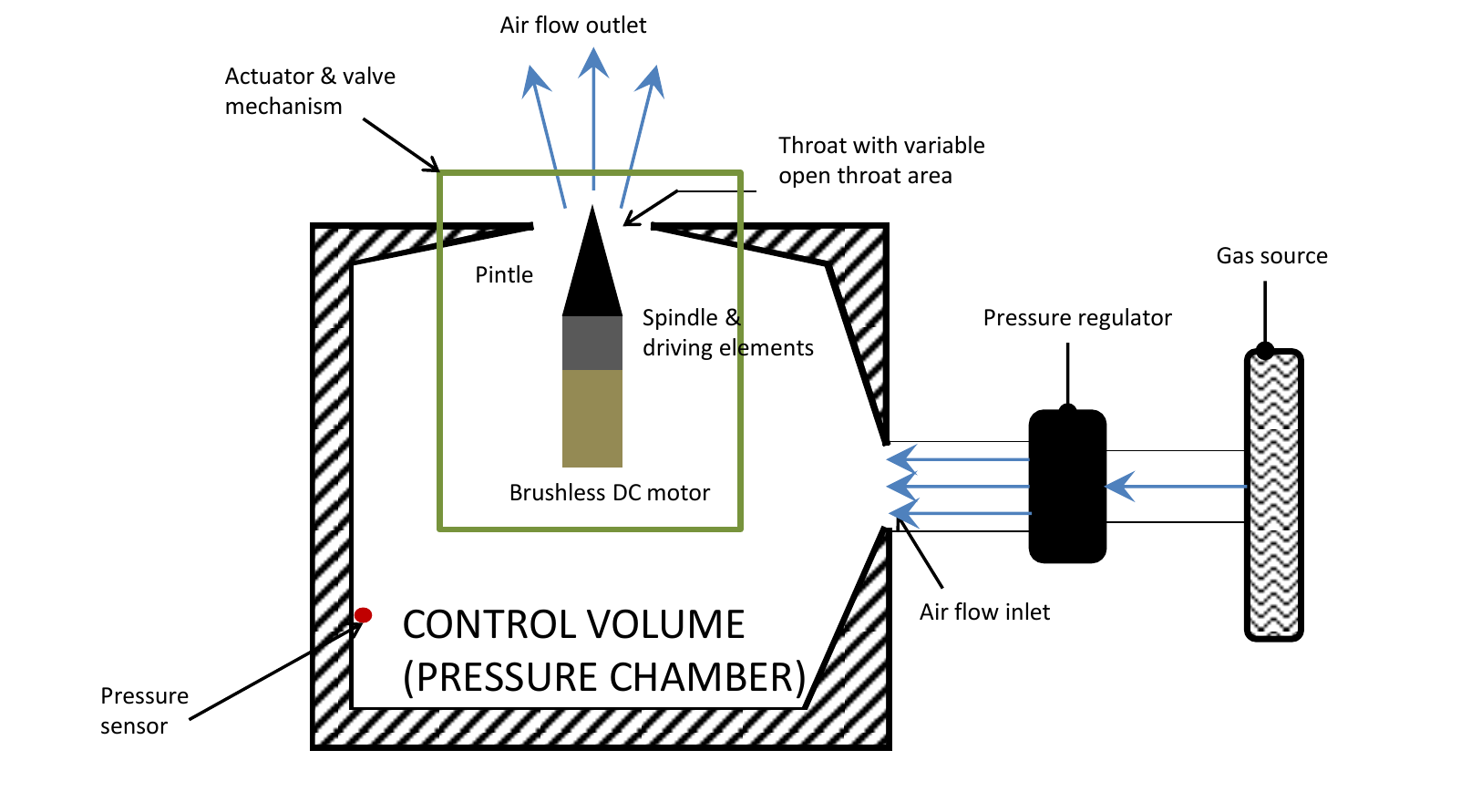}
	\caption{Schematic of cold air test setup}
	\label{fig:CATP}
\end{figure}

Overall cold air test setup (CATS) consists of a control volume (pressure chamber), an actuator, a valve mechanism, drive-train elements, a gas supply and a pressure regulator (see Fig. \ref{fig:CATP}). A continuous flow of gas is provided by a nitrogen source to the plant from the inlet and flow is adjusted by a pressure regulator. The output of the model, which is the pressure inside the control volume, is controlled through changing the exit throat area of the flow, which is the model input. The throat area is increased/decreased using the linear motion of a pintle at the exit throat. Drive-train elements are used to convert the rotational motion of the actuator, which is a brushless direct current motor, to translational motion of the pintle with required amount of reduction. 

\subsection{Plant} \label{sec:plant}

Assuming ideal gas conditions, the difference between the mass flow rates going into the control volume (pressure chamber), $\dot{m}_{in}$ [kg/sec], and going out of the control volume, $\dot{m}_{out}$ [kg/sec], is given as 
\begin{equation} \label{eq:perfectgas}
\dot{m}_{in}-\dot{m}_{out}=\frac{\dot{P}V}{RT}
\end{equation}
where $R$ [J/(kg K)] is the specific gas constant and $P$ [Pa], $T$ [K] and $V$ [m\textsuperscript{3}] are the pressure, the temperature and the volume of the gas inside the control volume. The process is assumed to be isothermal with no change in the control volume, hence $\dot{V}=\dot{T}=0$.

Mass flow coming out of the control volume is a function of the throat area ($A_t$ [mm\textsuperscript{2}]) and the pressure inside the control volume ($P$). Assuming chocked flow conditions, the relationship between the throat area and the resulting mass flow rate out of the control volume is given by \cite{Thomaier:87}
\begin{equation} \label{eq:massflowout}
\dot{m}_{out}=PA_t((\frac{2}{\gamma+1})^{(\frac{\gamma}{\gamma-1})})\sqrt{\frac{\gamma}{RT^*}}=\frac{PA_t}{c^*},
\end{equation}
where $\gamma$ is the specific heat ratio of air, $T^*$ is the temperature at the throat and $c^*$ [m/s] is the characteristic velocity of the gas inside the control volume. Using (\ref{eq:perfectgas}) and (\ref{eq:massflowout}), it is obtained that
\begin{equation} \label{eq:EOM2}
\dot{P}=c_1\dot{m}_{in}-c_2PA_t=M-c_2PA_t,
\end{equation}
where $c_1=\frac{RT}{V}$, $c_2=\frac{c_1}{c^*}$ and $M=c_1\dot{m}_{in}$. 

\subsection{Actuator} \label{sec:actuator}
A brushless DC motor is used in position controller mode as the actuator, with its driver card and an encoder to measure the position/speed of the rotor. The closed loop actuator dynamics is approximated as a first order linear system, whose transfer function is given as
\begin{equation}  \label{eq:actuator}
W_{act}(s)=\frac{\theta_{mes}(t)}{\theta_{com}(t )}=\frac{1}{\tau_{act}s+1}
\end{equation}
where $\theta_{mes}$ [quadrature] is the measured actual rotational position of the rotor while $\theta_{com}$ [quadrature] is the commanded rotational position (4000 quadratures (qc) correspond to 1 rotation) and $\tau_{act}$ is the actuator closed loop time constant.

\subsection{Valve geometry and drive-train elements} \label{sec:valvedrive}
\begin{figure} [htp]
	\centering
	\includegraphics[trim = 0mm 40mm 50mm 30mm, clip, width=1\textwidth]{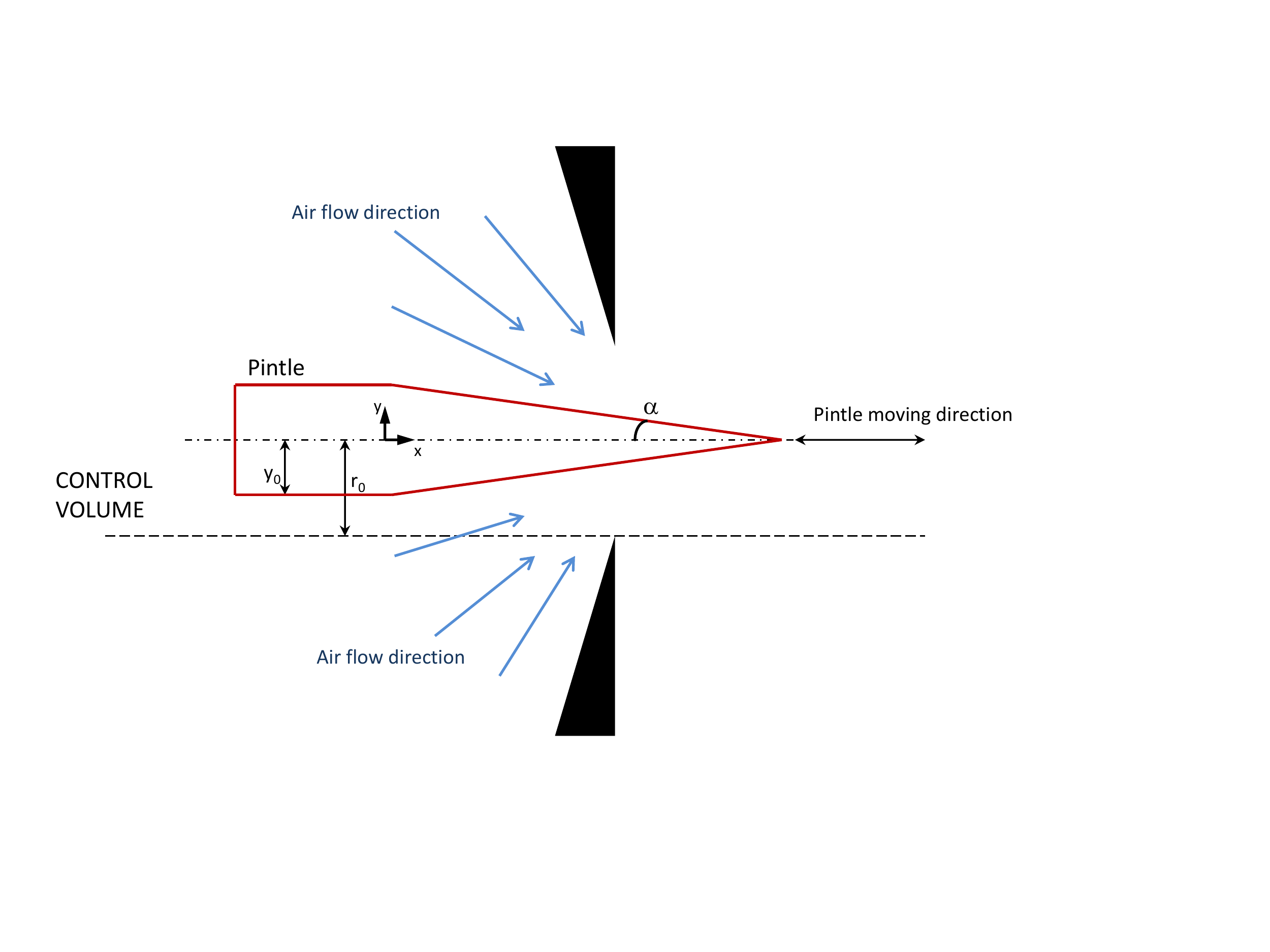}
	\caption{Valve geometry}
	\label{fig:valve}
\end{figure}

The drive-train elements consist of a gear box and a spindle to convert the rotational motion of the motor into the translational motion of a pintle at the throat area (see Fig. \ref{fig:valve}). The linear position of the pintle determines the throat opening.

The pintle has one degree of freedom in $x$ direction and open throat area changes as the pintle moves along the $x$ axis due to its conical surface. The cross sectional area of the cylindrical part at the back of the pintle is smaller than the fixed throat area, which makes sure that the open throat area $A_t$ is always larger than zero and protects the system from rapid pressure build up. Below, we provide the valve and the drive train models.

\subsubsection{Valve Geometry} \label{sec:valve}
There exist a nontrival  relationship between the movement of the pintle and the minimum throat area, where the choked flow conditions occur, due to their complex geometries \cite{Lee:13catp,Heo:15}. The size and the location of the minimum throat area is hard to estimate analytically due to the fact that location of the choked flow line, where the throat area is minimum, shifts towards the upstream as the pintle moves into the throat \cite{Heo:15}. In this study, the size of the open throat area is approximated as the projection of the real area on the vertical surface that is perpendicular to the pintle center line. Movement of the pintle along the $x$ axis reduces the projected throat area by
\begin{equation}   \label{eq:valve1}
y=y_0-tan(\alpha)x,
\end{equation}
where $y_0$ is the radius of the pintle at cylindrical part and $\alpha$ is the half of the cone angle at the tip of the pintle (see Fig. \ref{fig:valve}). The projected open throat area is then calculated as
\begin{equation}  \label{eq:valve2}
A_t=(r_0^2-y^2)\pi.
\end{equation}

\subsubsection{Drive-Train Elements} \label{sec:drive-train}
A gear box with a reduction ratio of $1:R_1$ is used to increase the torque output of the actuator. The spindle has an $R_2$ [mm] thread pitch, i.e. one turn of rotation corresponds to $R_2$ [mm] translational motion. Therefore, the relationship between the actuator rotational position, $\theta$ [quadrature], and the linear position of the pintle, $x$ [mm], can be calculated as
\begin{equation}  \label{eq:valve3}
x=\frac{\theta R_2}{R_1 \times 4000}
\end{equation}
where 4000 quadratures $(qc)$ correspond to 1 rotation.

Using (\ref{eq:valve1}-\ref{eq:valve3}), it is obtained that
\begin{equation}  \label{eq:valve}
A_t=(a_1+a_2 \ \theta + a_3 \ \theta^2) \pi
\end{equation}
where $a_1=r_0^2-y_0^2$, $a_2=\frac{2y_0tan(\alpha)R_2}{R_1 \times 4000}$ and $a_3=-(\frac{tan(\alpha)R_2}{R_1 \times 4000})^2$. 

\subsection{Model Enhancements Using Experimental Data} \label{sec:update}
To improve the fidelity of the system model, open loop experimental tests are performed and the obtained experimental data is used to adjust model parameters.
\begin{figure} [h]
	\centering
	\includegraphics[trim = 0mm 99mm 0mm 99mm, clip, width=1\textwidth]{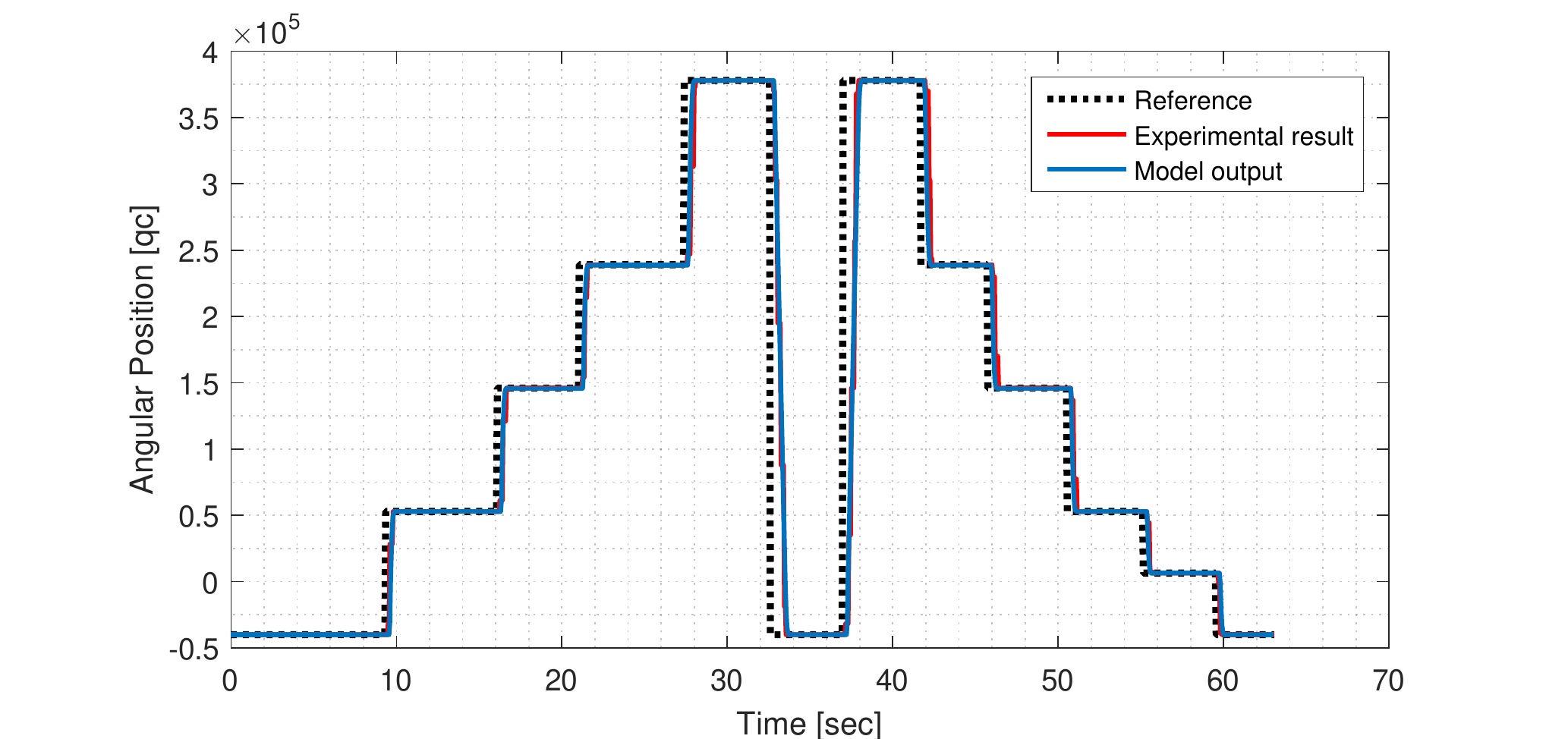}
	\caption{Open loop test results and updated model results of the actuator}
	\label{fig:Open1}
\end{figure}
Firstly, actuator model is updated: Brushless DC motor is commanded to track inputs in the position controller mode and based on the response of the actuator the time constant $\tau_{act}$ in (\ref{eq:actuator}) is updated. Experiments also revealed that a considerable amount of time delay exists in the actuator control loop, which is due to the communication and computation lags. After adjusting the time constant and incorporating a time delay, the enhanced actuator model output is compared with the experimental results and the outcomes are presented  in Fig. \ref{fig:Open1}, which shows that the updated model has a good agreement with the test data. 
\begin{figure} [h]
	\centering
	\includegraphics[trim = 0mm 99mm 0mm 99mm, clip, width=1\textwidth]{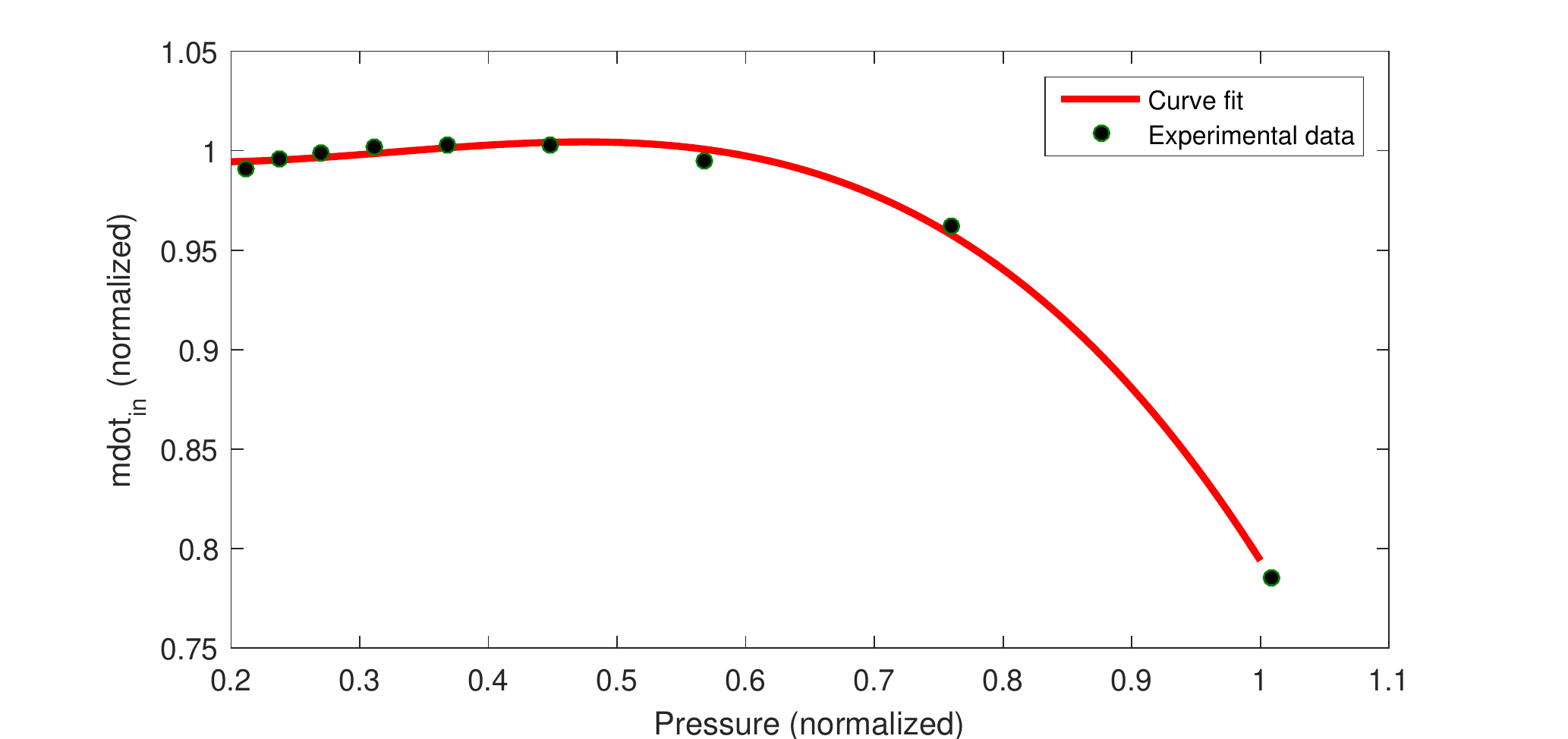}
	\caption{$\dot{m}_{in}$ calculated from (\ref{eq:EOM2}) in tests and curve fitted to the data (\ref{eq:EOM4})}
	\label{fig:Open2}
\end{figure}
To improve the system model further, parameters in (\ref{eq:EOM2}) are considered next: $R$, $T$, $V$ and $c^*$ are available for the test conditions with good accuracy, and therefore the values of these parameters are easily obtained. However mass flow rate ($\dot{m}_{in}$) is not always feasible to measure, especially for relatively small flow rate values. Therefore, the mass flow rate going into the CATS plant is calculated via (\ref{eq:EOM2}) using steady state pressure values at different operating points and corresponding throat areas. Several values for $\dot{m}_{in}$ at different operating points are plotted in Fig. \ref{fig:Open2} together with a polynomial fit. At low plant pressure, mass flow rates are nearly constant. However, mass flow rate decreases at higher pressure, because high back pressure overcomes the mechanical force in the pressure regulator and reduces the flow rate. Using the polynomial that is fitted to the data in Fig. \ref{fig:Open2}, (\ref{eq:EOM2}) is updated as
\begin{equation} \label{eq:EOM4}
\dot{P}=c_1(c_3P^3+c_4P^2+c_5P+c_6)-c_2PA_t.
\end{equation}

\begin{figure} [h]
	\centering
	\includegraphics[trim = 0mm 99mm 0mm 99mm, clip, width=1\textwidth]{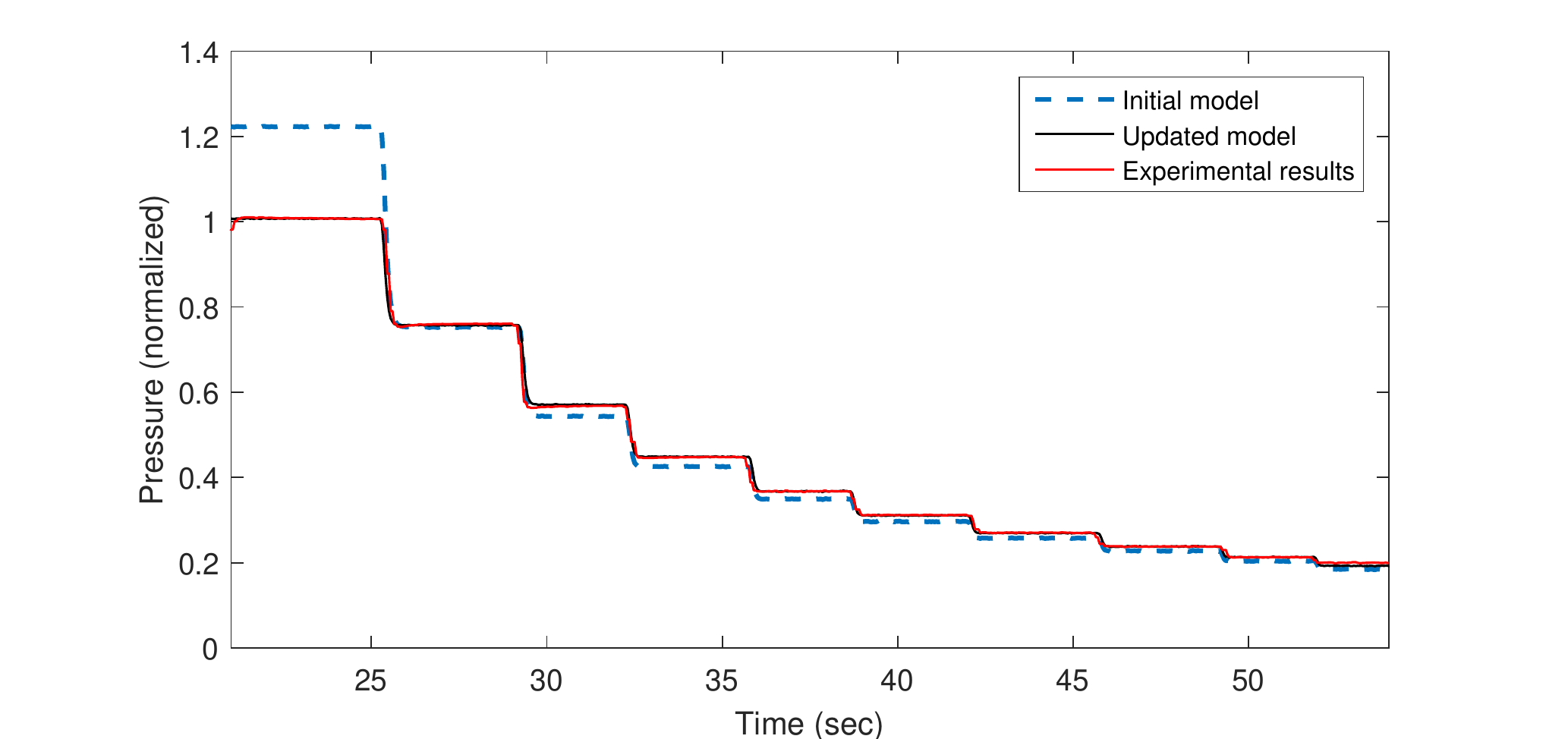}
	\caption{Open loop responses of experimental setup and simulation with initial plant model (\ref{eq:EOM2}) and updated plant model (\ref{eq:EOM4})}
	\label{fig:Open3}
\end{figure}
Open loop simulation results with the overall updated system model along with experimental results, which are obtained for a range of operating points, are given in Fig. \ref{fig:Open3}. It is noted that the model enhancements can be improved further by making comparisons at several other operating points followed by further tuning of the parameters but it is determined that this level of fidelity is enough for simulation evaluation purposes. 

\subsection{Modeling For Controller Design} \label{sec:simplification}

As explained earlier, the nonlinear model of the CATS developed in the earlier sections is used to evaluate controller alternatives in the simulation environment. To facilitate the controller design, a simpler model is developed following the steps listed below:
\begin{itemize}
	\item	Nonlinear plant model and valve equation in (\ref{eq:EOM2}) and (\ref{eq:valve}), respectively, are linearized.
	\item   Actuator dynamics are ignored due to small time constants compared to the plant.
\end{itemize}

Linearizing (\ref{eq:EOM2}) around an equilibrium point $(P,A_t)=(P_0,A_{t0})$, it is obtained that
\begin{equation} \label{eq:EOM3}
\dot{P}=\Delta \dot{P}= (-c_2A_{t0})\Delta P-(c_2P_0) \Delta A_t.
\end{equation}
where $\Delta P = P-P_0$ and $\Delta A_t = A_t-A_{t0}$. Defining $a_p \equiv -c_2A_{t0}$ and $b_p \equiv -c_2P_0$, (\ref{eq:EOM3}) can be rewritten as
\begin{equation} \label{eq:linearized}
\Delta \dot{P}= a_p\Delta P+b_p\Delta A_t.
\end{equation}
The value of $a_3$ in (\ref{eq:valve}) is much smaller than $a_1$ and $a_2$ for meaningful physical parameters, and therefore (\ref{eq:valve}) is approximated as
\begin{equation}  \label{eq:valve_lin}
A_t \approx (a_1+a_2 \ \theta) \pi.
\end{equation}
It is noted that the valve equation (\ref{eq:valve_lin}) is used to convert the required throat area determined by the pressure controller to the required actuator rotational position, which is provided to the actuator/valve control mechanism as a reference (see Fig. \ref{fig:thrustloop}).

Therefore, together with the actuator time lag, $\tau$, the system model used to develop the controller is obtained that
\begin{equation}  \label{eq:overall}
W_{p}(s)=\frac{\Delta P(t)}{\Delta A_{t,com}(t)}=\frac{b_p e^{-s\tau}}{s-a_p}
\end{equation}

\section{Controller Design} \label{sec:controller}
The structure of the closed loop control system of a conventional throttleable ducted rocket is provided in Fig. \ref{fig:thrustloop}. The outer loop determines the required gas generator (GG) pressure to obtain a desired thrust/speed profile, and the required pressure becomes the reference for the inner loop pressure controller. The pressure controller is the main focus of this paper and will be referred to as ``the controller" in the following sections. 4 different controllers are evaluated both in the simulation environment and in the experimental set-up: Model reference adaptive controller (MRAC), closed loop reference model (CRM) adaptive controller, the delay resistant closed loop reference model (DR-CRM) adaptive controller, which is introduced in this paper, and a PI controller. These controllers are explained below in the given order. It is noted that although MRAC is a very well known controller, its development is also provided here to explain the sequential controller improvement steps starting from MRAC and ending with the DR-CRM adaptive control. 

\subsection{Model reference adaptive controller (MRAC)} \label{sec:MRAC}
Consider following plant dynamics
\begin{equation}  \label{eq:MRAC1}
	y_p(t)=W_p(s)u(t), \ \ \ W_p(s)=\frac{k_pZ_p(s)}{R_p(s)},
\end{equation}
where $y_p \in \Re$ and $u \in \Re$ are the measured output and the control input of the system, respectively. $Z_p(s) \text{ and } R_p(s)$ are monic polynomials with orders of $m \text{ and } n$ and $k_p \in \Re$ is the constant gain of the plant.
Following assumptions are made for the plant \cite{Ann:05}:
\begin{itemize}
	\item System order $n$ is known along with the relative degree $n^*=n-m$.
	\item Sign of $k_p$ is known.
	\item Polynomial $Z_p(s)$ is Hurwitz.
\end{itemize}

The reference model, which gives the desired response of the closed loop system, is given as
\begin{equation}  \label{eq:MRAC2}
y_m(t)=W_m(s)r(t), \ \ \ W_m(s)=\frac{k_mZ_m(s)}{R_m(s)},
\end{equation}
where $y_m \in \Re$ and $r \in \Re$ are the output of the reference model and bounded reference signal, respectively. $W_m(s)$ is chosen as strictly positive real with relative degree equal to the relative degree of the plant.

State space description of the plant and the signal generators for the output feedback problem with controllable ($\Lambda,b_\lambda$) pair are given as
\begin{equation}  \label{eq:MRAC_ss}
\begin{split}
\dot{x}_p(t)& =A_px_p(t)+b_pu(t), \ \ \ \ y_p(t)=h_p^Tx_p(t)\\
\dot{\omega}_1(t)& =\Lambda \omega_1(t)+b_\lambda u(t) \\
\dot{\omega}_2(t)& =\Lambda \omega_2(t)+b_\lambda y_p(t)
\end{split}
\end{equation}
where $x_p \in \Re^{n}$, $\omega_1 \in \Re^{n-1}$, $\omega_2 \in \Re^{n-1}$, $A_p \in \Re^{n \times n}$, $b_p \in \Re^n$, $h_p \in \Re^n$, $\Lambda \in \Re ^{(n-1) \times (n-1)}$ is Hurwitz and $b_\lambda \in \Re ^{n-1}$.

It can be shown there exist constant parameters $\theta_0 \in \Re$, $\theta_1 \in \Re^{n-1}$, $\theta_2 \in \Re^{n-1}$ and $\theta_r \in \Re$ such that the controller given as 
\begin{equation}  \label{eq:MRAC3}
u(t)=\theta_0y_p(t)+\theta_1^T\omega_1(t)+\theta_2^T\omega_2(t)+\theta_{r}r(t)
\end{equation}
satisfies the desired reference model response characteristics \cite{Ann:05}. When the plant parameters are unknown, for $n^*=1$, the following adaptation law
\begin{equation}  \label{eq:MRAC6}
\dot{\Theta}(t) = -\Gamma \text{sign}(k_p) e_1(t) \Omega(t) 
\end{equation}
where 
\begin{equation}  \label{eq:MRAC4}
\Theta(t)=\begin{bmatrix}
\theta_0(t)  \\
\theta_1(t)  \\
\theta_2(t) \\
\theta_r(t)
\end{bmatrix},
\quad
\Omega(t)=\begin{bmatrix}
y_p(t)  \\
\omega_1(t)  \\
\omega_2(t) \\
r(t)
\end{bmatrix},
\end{equation}
$e_1(t)=y_p(t)-y_m(t)$ is the tracking error, $\Gamma \in \Re^{2n \times 2n} $ is a diagonal matrix with positive elements, stabilizes the closed loop system and ensures that $e_1 \rightarrow 0$ as $t \rightarrow \infty$ \cite{Ann:05}. 

\subsection{Closed-loop reference model (CRM) adaptive controller} \label{sec:CRM}
The reference model in the classical model reference adaptive control (\ref{eq:MRAC2}) is unaffected by the tracking error. In CRM adaptive controller, however, the tracking error $e_1(t)=y_p(t)-y_m(t)$ is fed back to the reference model for the purpose of improving the transient dynamics \cite{Gibson:12}. Consider the following state space representation of the reference model dynamics given in (\ref{eq:MRAC2})
\begin{equation}  \label{eq:ref_ss}
\dot{x}_m(t) =A_mx_m(t)+b_mr(t), \ \ \ \ y_m(t)=h_m^Tx_m(t),
\end{equation}
where $A_m \in \Re^{n \times n}$, $b_m \in \Re^{n}$ and $h_m \in \Re ^{n}$. In classical model reference adaptive control, $A_m,b_m$ and $h_m$ are chosen such that the transfer function $h_m^T(sI-A_m)b_m=W_m(s)=k_m\frac{Z_m(s)}{R_m(s)}$ becomes strictly positive real. In CRM adaptive controller, the reference model is modified as
\begin{equation}  \label{eq:CRM1}
\dot{x}_m(t)=A_mx_m(t)+b_mr(t)+L(y_p(t)-y_m(t)), \ \ \ \ y_m(t)=h_m^Tx_m(t),
\end{equation}
where $L \in \Re^n$ is a design parameter vector. The relationship between the reference model output $y_m$, the reference $r$ and the tracking error $e_1$ then becomes
\begin{equation}  \label{eq:CRM2}
y_m(t)=W_m(s)r(t)+W_L(s)e_1(t)
\end{equation}
where 
\begin{equation}
h_{m}^T(sI-A_{m})L=k_L\frac{Z_L(s)}{R_m(s)}=W_L(s).
\end{equation}
Polynomial $Z_L(s)$ is order of $n-1$. Boundedness of all the signals in the closed loop system along with the convergence of the tracking error as in classical MRAC is valid for the CRM approach, using the same controller structure (\ref{eq:MRAC3}) and the adaptive law (\ref{eq:MRAC6}), as long as $L$ is chosen such that the transfer function
\begin{equation} \label{eq:SPR}
W_e=\frac{Z_m(s)}{R_m(s)-k_LZ_L(s)}
\end{equation}
is strictly positive real (see \cite{Gibson:12_2}).

\subsection{Delay resistant closed loop reference model (DR-CRM) adaptive controller} \label{sec:DR-CRM}
Consider the following plant with an input time delay
\begin{equation} \label{eq:plant}
y_p(t)=k_p\frac{Z_p(s)}{R_p(s)}u(t-\tau)=W_p(s)u(t-\tau)
\end{equation}
where $y_p \in \Re$ is the measured output, $u \in \Re$ is the control signal, $\tau$ is the known time delay, $Z_p(s)$ and $R_p(s)$ are monic coprime polynomials with orders of $m$ and $n$, respectively, and $k_p \in \Re$ is the constant gain of the plant. Following assumptions are made for the plant:
\begin{itemize}
	\item System order $n$ is known along with the relative degree $n^*=n-m$.
	\item Sign of $k_p$ is known.
	\item Polynomial $Z_p(s)$ is Hurwitz.
\end{itemize}

The reference model dynamics are given with the closed loop reference model structure as
\begin{equation} \label{eq:ref}
\begin{split}
\dot{x}_m(t)& = A_{m}x_m(t)+b_{m}r(t-\tau)+L(y_p(t)-y_m(t)) \\
y_m(t) & = h_{m}^T x_m(t) 
\end{split}
\end{equation}
where $x_m \in \Re^n$, $y_m \in \Re$, $r \in \Re$, $A_m \in \Re^{n \times n}$ and $b_m,L,h_m \in \Re^n$. Input-output relationship of the closed loop reference model is given as
\begin{equation} \label{eq:ym}
y_m(t)=W_m(s)r(t-\tau)+W_L(s)e_1(t)
\end{equation}
where $e_1=y_p-y_m$ is the tracking error. The transfer functions describing the closed loop reference model are
\begin{equation}
\begin{gathered}
	h_{m}^T(sI-A_{m})b_m=k_m\frac{Z_m(s)}{R_m(s)}=W_m(s) \\
	h_{m}^T(sI-A_{m})L=k_L\frac{Z_L(s)}{R_m(s)}=W_L(s)
\end{gathered}
\end{equation}
where $R_m(s)$ is a monic polynomial with order $n$ while $Z_m(s)$ and $Z_L(s)$ are two monic polynomials with order $n-1$, $k_m \in \Re$ and $k_L \in \Re$ are the gains of the transfer functions.

It is noted that under model matching conditions the tracking error becomes zero, which reduces the reference model (\ref{eq:ym}) to 
\begin{equation} \label{eq:ym2}
y_m(t)=W_m(s)r(t-\tau).
\end{equation}

State space description of the plant (\ref{eq:plant}) and the signal generators for the output feedback problem with the controllable ($F,g$) pair are given as
\begin{equation} \label{eq:plant_ss}
\begin{split}
\dot{x}_p(t)& =A_px_p(t)+b_pu(t-\tau), \ \ \ y_p(t)=h_p^Tx(t) \\
\dot{\omega}_1(t)& =F \omega_1(t)+g u(t-\tau) \\
\dot{\omega}_2(t)& =F \omega_2(t)+g y_p(t)
\end{split}
\end{equation}
where $x_p \in \Re^n$ is the state vector, $\omega_1,\omega_2 \in \Re^n$, $A_p \in \Re ^{n \times n}$, $b_p$, $h_p \in \Re ^n$, $F \in \Re ^{n \times n}$ is Hurwitz and $g \in \Re ^n$. It can be shown that there exists constant controller parameters $\beta_1^* \in \Re^n$, $\beta_2^*  \in \Re^n$ and $k^* \in \Re$ such that the controller
\begin{equation} \label{eq:u_matching}
u(t) = \beta_1^{*T}\bar{\omega}_1(t) + \beta_2^{*T}\bar{\omega}_2(t) + k^*r(t)
\end{equation}
where $\bar{\omega}_1(t) \triangleq \omega_1(t+\tau)$, $\bar{\omega}_2(t) \triangleq \omega_2(t+\tau)$ together with the rewritten plant dynamics 
\begin{equation} \label{eq:plant_ss_noncas}
\begin{split}
\dot{\bar{x}}_p(t)& =A_p\bar{x}_p(t)+b_pu(t), \ \ \ \bar{y}_p(t)=h_p^T\bar{x}_p(t) \\
\dot{\bar{\omega}}_1(t)& =F \bar{\omega}_1(t)+g u(t) \\
\dot{\bar{\omega}}_2(t)& =F \bar{\omega}_2(t)+g \bar{y}_p(t)
\end{split}
\end{equation}
where $\bar{x}_p(t) \triangleq x_p(t+\tau)$ and $\bar{y}_p(t) \triangleq y_p(t+\tau)$, satisfy the model matching conditions \cite{Yil:Auto}. 

It is shown in \cite{Ann:05} that the plant output, $y_p(t)$, can be expressed as a linear combination of $\omega_1(t),\omega_2(t)$ as
\begin{equation} \label{eq:yw1w2}
y_p(t)=c^T \omega_1(t) + d^T \omega_2(t)
\end{equation}
where $c,d \in \Re^n$. When (\ref{eq:yw1w2}) is substituted into (\ref{eq:plant_ss}), it is obtained that 
\begin{equation} \label{eq:w1w2}
\begin{bmatrix}
\dot{\omega}_1(t)  \\
\dot{\omega}_2(t)
\end{bmatrix}  = 
A
\begin{bmatrix}
\omega_1(t)  \\
\omega_2(t)
\end{bmatrix}
+bu(t-\tau),
\end{equation}
where $A \in \Re^{2n \times 2n}$ and $b \in \Re^{2n}$ are given as $A=\begin{bmatrix}
F & 0 \\ gc^T & F+gd^T
\end{bmatrix}$ and $b=\begin{bmatrix}
g \\ 0
\end{bmatrix}$.
Non-casual terms in (\ref{eq:u_matching}) can then be calculated as
\begin{equation} \label{eq:future}
\begin{bmatrix}
\bar{\omega}_1(t)  \\
\bar{\omega}_2(t)
\end{bmatrix}  = 
e^{A\tau}
\begin{bmatrix}
\omega_1(t)  \\
\omega_2(t)
\end{bmatrix}
+\int_{-\tau}^{0}e^{A\eta}bu(t+\eta)d\eta.
\end{equation}
When (\ref{eq:future}) is substituted into (\ref{eq:u_matching}), the control signal becomes
\begin{equation} \label{eq:u_DRCRM}
u(t)=\alpha^{*T}_1\omega_1(t)+\alpha^{*T}_2\omega_2(t)+\int_{-\tau}^{0}\phi^*(\eta)u(t+\eta)d\eta +k^*r(t)
\end{equation}
where $\alpha^*_{1,2} \in \Re^{n}$, $\phi^*(\eta) \in \Re$ are the corresponding controller parameters which eliminate the non-causality in controller (\ref{eq:u_matching}) using (\ref{eq:future}). 

In the case of unknown plant parameters, the control input, $u(t)$, can be split into two sub-signals as
\begin{equation} \label{eq:u=u1+u2}
 u(t) = u_1(t) + u_2(t)
\end{equation}
where
\begin{equation} \label{eq:u1}
u_1(t)=\alpha^{*T}_1\omega_1(t)+\alpha^{*T}_2\omega_2(t)+\int_{-\tau}^{0}\phi^*(\eta)u(t+\eta)d\eta +k^*r(t)
\end{equation}
and
\begin{equation} \label{eq:u2}
u_2(t)=\tilde{\alpha}^{T}_1(t)\omega_1(t)+\tilde{\alpha}^{T}_2(t)\omega_2(t)+\int_{-\tau}^{0}\tilde{\phi}(t,\eta)u(t+\eta)d\eta +\tilde{k}(t)r(t)
\end{equation}
where 
$\tilde{\alpha}_{i}(t)=\alpha_{i}(t)-\alpha^*_{i} \text{ for } i=1,2; \ \tilde{\phi}(t,\eta)=\phi(t,\eta)-\phi^*(\eta)$ and $\tilde{k}(t)=k(t)-k^*$.

Substituting (\ref{eq:u=u1+u2}), using (\ref{eq:u1}) and (\ref{eq:u2}), into (\ref{eq:plant_ss}), the closed loop dynamics is obtained as
\begin{equation} \label{eq:X_p}
\begin{split}
\dot{X}_p(t)&=A_{mn}X_p(t)+b_{mn} \Bigg[\tilde{\theta}(t-\tau)w(t-\tau)+\int_{-\tau}^{0}\tilde{\phi}(t-\tau,\eta)u(t-\tau+\eta)d\eta+k^*r(t-\tau) \Bigg] \\
y_p(t) & =h_{mn}^T X_p(t) 
\end{split}
\end{equation}
where 
$
	A_{mn} = \begin{bmatrix}
		A_p & b_p\beta^*_1 & b_p\beta^*_2 \\
		0 & F+g\beta^*_1 & g\beta^*_2 \\
		g h_p^T & 0 & F
	\end{bmatrix}, \ \ \
	b_{mn} = \begin{bmatrix}
		b_p  \\
		g  \\
		0
	\end{bmatrix}, \ \ \
	h^T_{mn} = \begin{bmatrix}
		h_p^T & \ \ 0 & \ \ 0
	\end{bmatrix}, \\
	X_p(t) = \begin{bmatrix}
		x^T_p(t) \ \ \omega^T_1(t) \ \ \omega^T_2(t)
	\end{bmatrix}^T, \
	w(t) = \begin{bmatrix}
		\omega^T_1(t) \ \  \omega^T_2(t) \ \ r(t)
	\end{bmatrix}^T  \ \text{and} \ \
	\tilde{\theta}(t) = \begin{bmatrix}
		\tilde{\alpha}_1(t) \ \ \tilde{\alpha}_2(t) \ \ \tilde{k}(t)
	\end{bmatrix}.
$

It is noted that when the parameter errors are zero, i.e. $\tilde{(\cdot)} = 0$, the closed loop dynamics represented by (\ref{eq:X_p}) becomes equivalent to that of the reference model dynamics (\ref{eq:ym2}), which shows that the system formed by ($A_{mn}, b_{mn}, h_{mn}$) is a non-minimal representation of the reference model:
\begin{equation} \label{eq:Wmn}
	h_{mn}^T(sI-A_{mn})b_{mn} \equiv k_p\frac{Z_m(s)}{R_m(s)}=\frac{k_p}{k_m}W_m(s).
\end{equation}

Using (\ref{eq:X_p}) and (\ref{eq:Wmn}), the plant output is obtained as
\begin{equation} \label{eq:yp}
y_p(t)=\frac{k_p}{k_m}W_m(s)\Bigg[\tilde{\theta}(t-\tau)w(t-\tau)+\int_{-\tau}^{0}\tilde{\phi}(t-\tau,\eta)u(t-\tau+\eta)d\eta+k^*r(t-\tau)\Bigg].
\end{equation}
Subtracting (\ref{eq:ym}) from (\ref{eq:yp}), it can be obtained as
\begin{equation} \label{eq:e11}
e_1(t)=\frac{k_p}{k_m}W_m(s)\Bigg[\tilde{\theta}(t-\tau)w(t-\tau)+\int_{-\tau}^{0}\tilde{\phi}(t-\tau,\eta)u(t-\tau+\eta)d\eta \Bigg] -W_Le_1(t).
\end{equation}
Solving (\ref{eq:e11}), the tracking error can be found as
\begin{equation} \label{eq:e1}
e_1(t)=k_p \frac{Z_m(s)}{R_m(s)-k_LZ_L(s)} \Bigg[\tilde{\theta}(t-\tau)w(t-\tau)+\int_{-\tau}^{0}\tilde{\phi}(t-\tau,\eta)u(t-\tau+\eta)d\eta \Bigg]
\end{equation}
where $\frac{Z_m(s)}{R_m(s)-k_LZ_L(s)}=W_e(s)$ has sufficient degrees of freedom, in terms of the design parameter vector $L$, to be determined as a SPR transfer function (see Section \ref{sec:CRM}).

The closed loop reference model in (\ref{eq:ref}) can be rewritten as,
\begin{equation} \label{eq:xm}
\begin{split}
\dot{x}_{mn}(t)& = A_{mn}x_{mn}(t)+b_{mn} k^*r(t-\tau) +GL(y_p(t)-y_m(t)) \\
y_m(t)& = h_{mn}^T x_{mn}(t)
\end{split}
\end{equation}
for $x_{mn}=\begin{bmatrix}x_p^{*T}(t) & \omega_1^{*T}(t) & \omega_2^{*T}(t) \end{bmatrix}^T$ where $x_p^{*}(t), \ \omega_1^{*}(t) \text{ and } \omega_2^{*}(t)$ are the signals in the reference model, corresponding to the signals $x_p(t), \ \omega_1(t) \text{ and } \omega_2(t)$ in the closed loop dynamics, respectively. $G \in \Re^{3n \times n}$ is the constant matrix to transform $x_m$ to the controllable subspace in $x_{mn}$ (see \cite{Gibson:12_2}). Error dynamics, $e(t)=X_p(t)-x_{mn}(t)$, in non-minimal form, is found by subtracting (\ref{eq:xm}) from (\ref{eq:X_p}) as
\begin{equation} \label{eq:E}
\begin{split}
\dot{e}(t)& =A_{e}e(t) +b_{mn} \Bigg[\tilde{\theta}(t-\tau) w(t-\tau) + \int_{-\tau}^{0} \tilde{\phi}(t-\tau,\eta) u(t-\tau+\eta) d\eta \Bigg] \\
\quad & e_1(t)=h_{mn}^T e(t)
\end{split}
\end{equation}
where
\begin{equation} \label{eq:Ae}
A_e=A_{mn}-GLh_{mn}^T.
\end{equation}

It can be shown \cite{Yil:Auto}, utilizing the error dynamics (\ref{eq:E}),
that the controller (\ref{eq:u=u1+u2})-(\ref{eq:u2}), along with the adaptation laws given as
\begin{equation} \label{eq:adaptations}
\begin{split}
\dot{\tilde{\theta}}(t) &=\dot{\theta}(t)=-\text{sign}(k_p) \Gamma_{\theta} e_1(t) \omega(t-\tau) \\
\dot{\tilde{\phi}}(t,\eta) &=\dot{\phi}(t)=-\text{sign}(k_p) \Gamma_{\phi} e_1(t) u(t-\tau+\eta) \quad -\tau \le \eta \le 0
\end{split}
\end{equation}
stabilizes the closed loop system as long as $L$ is chosen to ensure that $W_e(s)$ is SPR. Furthermore, tracking error $e_1(t)$ converges to zero. $\Gamma_\theta \in \Re^{(2n+1) \times (2n+1)}$ is a diagonal matrix with positive elements and $\Gamma_\phi \in \Re^+$.

\subsection{Proportional-integral controller (PI)} \label{sec:PI}
\begin{figure}[htp]
	\captionsetup[subfloat]{labelformat=empty}
	\centering
	\subfloat[h][]{\includegraphics[trim = 0mm 99mm 0mm 99mm, clip, width=12cm]{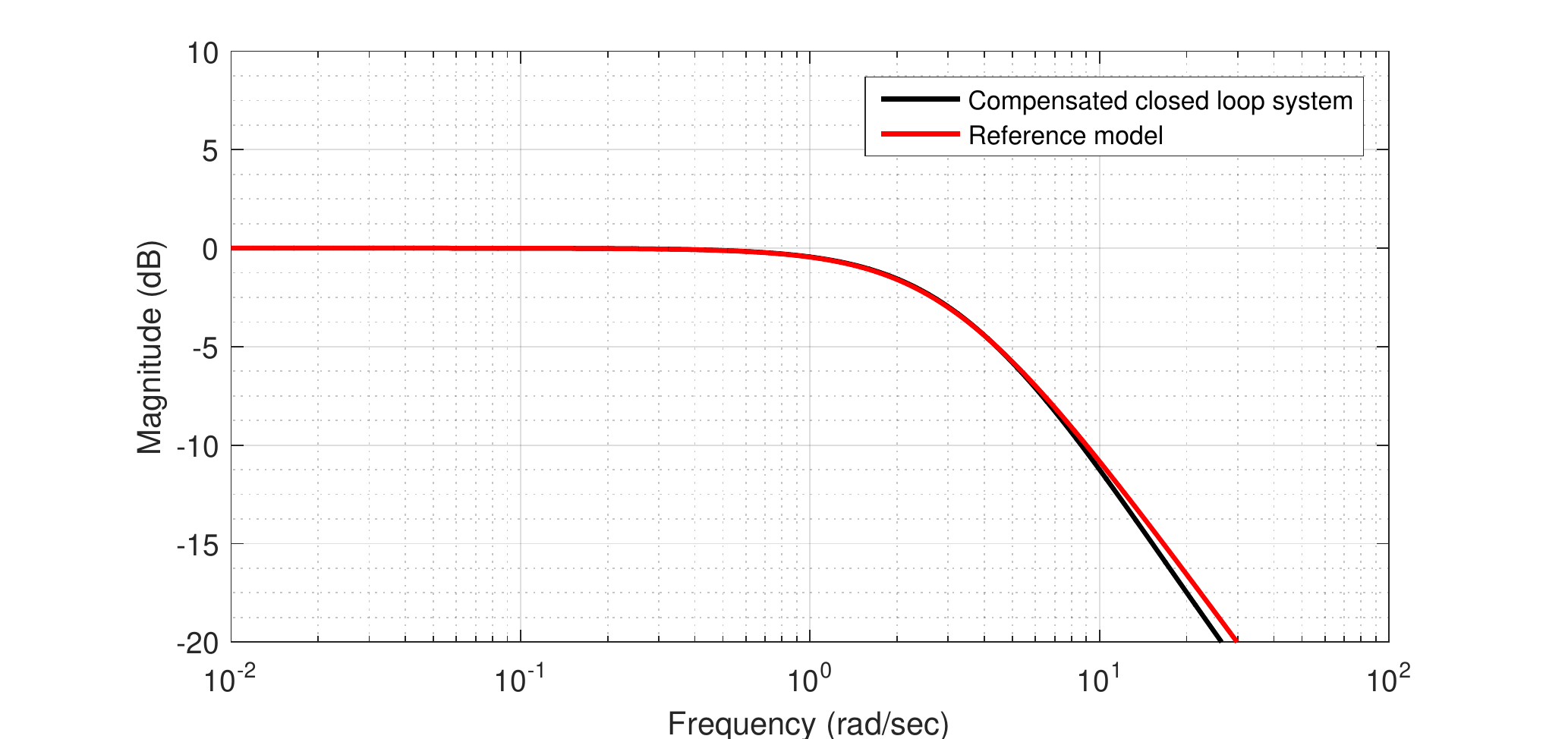}}
	
	\subfloat[h][]{\includegraphics[trim = 0mm 99mm 0mm 99mm, clip, width=12cm]{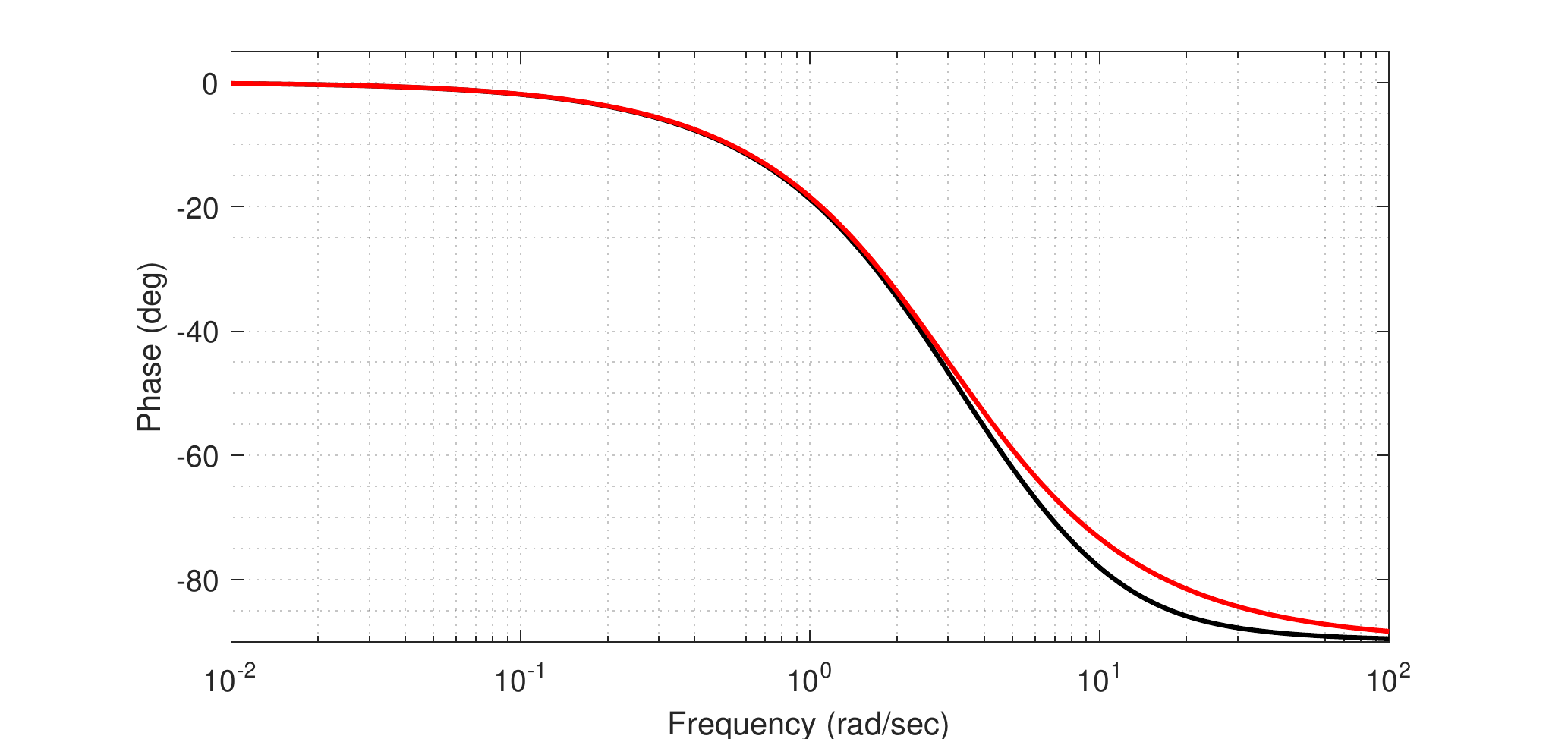}}
	
	\caption{Comparison of the Bode plots of the compensated closed loop system with the designed PI controller and the reference model}
	\label{fig:FR}
\end{figure}

A proportional-integral controller is designed for the plant dynamics given in (\ref{eq:MRAC1}). The controller transfer function is given by
\begin{equation}  \label{eq:PI1}
G_{PI}(s)=K_{p}(1+\frac{1}{T_i s})
\end{equation}
where scalar controller parameters $K_{p}$ and $T_i$ are selected to make the closed loop dynamics provide a similar response with the reference model (\ref{eq:MRAC2}) used for MRAC. Frequency response plots of the reference model and the compensated system are provided in Fig. \ref{fig:FR}.

\subsection{Implementation Enhancements} \label{sec:imp}
The implementation of the adaptive controllers requires some modifications to address the issues that were not taken into consideration during the initial design but arise in real experimental tests. Below we explain these experimental requirements and how they are addressed. 

\subsubsection{Disturbance rejection} \label{sec:firstorder}
The goal of the model reference adaptive control designs presented in Sections \ref{sec:MRAC}, \ref{sec:CRM}, \ref{sec:DR-CRM} is to force the plant output follow the reference model output while keeping all the system signals bounded. The disturbances, on the other hand, are not explicitly taken into account. In throttleable ducted rocket propulsion systems, the solid propellant in the gas generator contains metallic particles (see Fig. \ref{fig:TDR}), which can cause deposition or ablation at the throat between the gas generator and the ram combuster, which acts like an additive disturbance on the effective throat area. 
Therefore, it is important to incorporate disturbance rejection capabilities into the gas generator pressure controller. Since the plant model (\ref{eq:overall}) developed for the controller design is first order, the disturbance rejection modification explained below utilizes a scalar plant model, which simplifies the process.

Consider the plant
\begin{equation}  \label{eq:dist_plant}
y_p(t) = a_p y_p(t) + b_p (u(t) + d_0)
\end{equation}
where $y_p$ is the system output, $d_0$ is the unknown constant disturbance, $u$ is the plant input and $a_p$ and $b_p$ are the plant parameters.
An additional adaptive controller parameter, $\theta_3(t)$, is introduced to the controller structure for MRAC and CRM adaptive controller, which results in the following control signal
\begin{equation}  \label{eq:u_firstorder1}
	u(t) = \theta_0 (t) y_p (t) + \theta_r (t) r (t) + \theta_{3}(t)
\end{equation}
where $theta_0$, $theta_r$ and $\theta_3$ are the adaptive control parameters to be determined and $r$ is a bounded reference signal. The adaptation laws for the controller parameters are given as
\begin{equation} \label{eq:adaptivelaws_firstorder1}
\dot{\bar{\Theta}} (t) = -\text{sign}(b_p) \bar{\Gamma} e_1(t) \bar{\Omega}(t)
\end{equation}
where 
$
\bar{\Theta}^T(t)=\begin{bmatrix}
\theta_0(t) \ \ \theta_r(t) \ \ \theta_3(t)
\end{bmatrix}
$, 
$
\bar{\Omega}^T=\begin{bmatrix}
y_p(t) \ \ r(t) \ \ 1
\end{bmatrix}
$, $\bar{\Gamma} > 0$ is the diagonal adaptation rate matrix and $e_1$ is the tracking error given by $e_1=y_p-y_m$.
The reference model for the MRAC is appointed as
\begin{equation}  \label{eq:u_ref11}
\dot{y}_m (t) = a_m y_m (t) + b_m r(t),
\end{equation}
and the reference model for CRM adaptive controller is selected as,
\begin{equation}  \label{eq:u_ref12}
\dot{y}_m (t) = a_m y_m (t) + b_m r(t) + l(y_p(t)-y_m(t)),
\end{equation}
where $a_m,b_m,l \in \Re$, $a_m = -b_m$ for unity DC gain and $a_m < 0$ and $l > 0$. It is noted that for a first order plant, choosing $l>0$ satisfies the stability conditions.

Similarly, DR-CRM adaptive control signal is modified as
\begin{equation}  \label{eq:u_firstorder2}
u(t) = \alpha_y (t) y_p (t) + \int_{-\tau}^{0} \lambda(t,\eta)u(t+\eta) d\eta + k (t) r (t) + \theta_3 (t).
\end{equation}
The adaptive laws for the controller parameters $\alpha_y, \lambda, k, \theta_3 \in \Re $ are given as
\begin{equation} \label{eq:adaptivelaws_firstorder2}
\begin{split}
\dot{\theta} (t) &= -\text{sign}(b_p) \Gamma_{\theta} e_1(t) \omega(t - \tau) \\
\frac{\partial \lambda}{\partial t} (t,\eta) & = -\text{sign}(b_p) \gamma_{\lambda} e_1(t)u(t+\eta-\tau) \quad -\tau \le \eta \le 0
\end{split}
\end{equation}
where 
$
\theta^T(t)=\begin{bmatrix}
\alpha_y(t) \ \ k(t) \ \ \theta_3(t)
\end{bmatrix}
$, 
$
\omega^T(t)=\begin{bmatrix}
y_p(t) \ \ r(t) \ \ 1
\end{bmatrix}
$, $\Gamma_{\theta} > 0$ and $\gamma_{\lambda} \in \Re^+$ are adaptation rate constants.

The reference model for the DR-CRM adaptive controller is determined as
\begin{equation}  \label{eq:u_ref2}
\dot{y}_m (t) = a_m y_m (t) + b_m r(t-\tau) + l(y_p(t)-y_m(t)),
\end{equation}
where $a_m,b_m,l \in \Re$, $a_m = -b_m$ for unity DC gain and $a_m < 0$ and $l > 0$.

\subsubsection{Robustness} \label{sec:projection}
The design of the DR-CRM adaptive controller presented in Section \ref{sec:DR-CRM} portrayed an idealized case, where the delay free part of the plant dynamics are assumed to be linear and time-invariant with unknown but constant parameters. Furthermore, the measurements are assumed to be perfect. However, it is known that in reality, no plant is truly linear or finite dimensional. Parameters may vary with time and operating conditions, and measurements are always contaminated with noise. The plant model used for the controller design is always an approximation of reality. Therefore, we need a robustfying modification against possible parameter drifts in adaptive controller parameters, due to these non-ideal situations. One common remedy utilized to prevent parameter drift is the projection algorithm \cite{Lavretsky:book}, which is explained below. 

Consider a controller parameter vector $\Theta(t) \in \Re^k$. The upper bound on the controller parameter vector norm $||\Theta||$, to activate the projection modification is defined as $\Theta_{max} \in \Re^+$. A continuous and differentiable convex function $f: \Re^k \rightarrow \Re$ is introduced as 
\begin{equation}  \label{eq:projection_f}
f (\Theta) = \frac{||\Theta||^2 - \Theta_{max}^2}{\epsilon \Theta_{max}^2}
\end{equation}
where $\epsilon > 0 $ is a constant which defines the projection tolerance region where $||\Theta|| \leq \Theta_{max}(\sqrt{1+\epsilon})$ forms a hard bound on the parameter norm. The projection operator, $\text{Proj}:\Re^k \times \Re^k \rightarrow \Re^k$, is described as
\begin{equation}  \label{eq:projectionoperator}
\text{Proj}(\Theta,y) \triangleq \begin{cases}
y-\frac{\nabla f(\Theta) (\nabla f(\Theta))^T }{ ||\nabla f(\Theta)||^2 }yf(\Theta), & \text{if} \ \ ||\Theta||>\Theta_{max} \ \wedge \ y^T \nabla f(\Theta) > 0 \\
y, & \text{otherwise}
\end{cases}
\end{equation}
where $y \in \Re^k$, $\nabla f(\Theta) \in \Re^{k}$ is the gradient vector of $f$, evaluated at $\Theta$. 

The adaptive control law is then modified as
\begin{equation}  \label{eq:projection_adaptivelaws}
\dot{\Theta} = \text{Proj}(\Theta,-\text{sign}(k_p) \Gamma e_1\Omega)
\end{equation}
where $\Gamma \in \Re^{k \times k}$ is a diagonal matrix with positive elements, $e_1 \in \Re$ is the tracking error, $\Omega \in \Re^k$ is the regressor vector containing system signals. It can be shown that with the adaptive law (\ref{eq:projection_adaptivelaws}) utilizing projection, the parameter drift is prevented  \cite{Lavretsky:book}. 

Although we observed in the experiments that the introduction of the projection algorithm prevents parameter drift, there are no well defined procedures to determine the upper bound $\Theta_{max}$ on the controller parameter vector $\Theta$. One method is to calculate this bound using the worst case uncertainty case. Another method is to conduct several experiments without projection and observe the variation of controller parameters, which can help define a reasonable upper bound. A third approach is setting initial values for the controller parameters that would satisfy the matching conditions for the nominal plant dynamics and then determining the upper bound for the parameters as a certain percentage higher than these initial values. In our experiments, we utilized the second method. 

\subsubsection{Digital implementation of the integral term in DR-CRM adaptive controller} \label{sec:digital}
For computer implementation purposes, the finite integral term in (\ref{eq:u_firstorder2}) is approximated as,
\begin{equation} \label{eq:intapproximation}
\int_{-\tau}^{0} \lambda(t,\eta) u(t+\eta) d\eta = \sum_{i=1}^{m} \lambda_i(t)  u(t-i dt) = \bar{\lambda}^{T}(t) \bar{u}(t)
\end{equation}
where $dt = 50$ ms is the sampling interval of the implemented software, $\tau = 300$ ms is the time delay and $m=\frac{\tau}{dt}=6$. $\bar{\lambda} \in \Re^m$ is the vector containing parameters 
$
\bar{\lambda}^T(t) = \begin{bmatrix}
\lambda_1(t)  \ldots \lambda_m (t) 
\end{bmatrix}
$
and $\bar{u} \in \Re^m$ is the delayed input vector 
$
\bar{u}^T(t) = \begin{bmatrix}
u(t-dt)  \ldots u (t-mdt) 
\end{bmatrix}.
$

Adaptation law (\ref{eq:adaptivelaws_firstorder2}) is updated as
\begin{equation} \label{eq:discretized_laws}
\dot{\bar{\theta}}(t)=-\text{sign}(b_p) \bar{\Gamma}_{\bar{\theta}} \ e_1(t) \bar{\omega}(t)
\end{equation} \
where 
\begin{equation} \label{eq:discretized_vector}
\bar{\theta}(t)=\begin{bmatrix}
\alpha_y(t) \\ \lambda_1(t) \\ \vdots \\ \lambda_m(t) \\ k(t) \\ \theta_3(t)
\end{bmatrix} \ \
\bar{\omega}(t)=\begin{bmatrix}
y_p(t) \\ u(t-dt) \\ \vdots \\  u(t-mdt)  \\  r(t) \\ 1
\end{bmatrix}
\end{equation}
and $\bar{\Gamma}_{\bar{\theta}} > 0$ is a diagonal adaptation rate matrix.

\subsubsection{Initialization of the controller parameters} \label{sec:init}
One way to initialize the adaptive controller is to use initial values that would satisfy model matching for the nominal plant dynamics. However, in MRAC and CRM adaptive controller designs, initial parameters $\theta_0 (0)$ and $\theta_r (0)$ had to be lowered to obtain the best performance since the presence of the time delay prevents exact model matching due to the lack of exact delay compensation terms. 
On the other hand, the initial parameters for DR-CRM adaptive controller are set to satisfy the model matching conditions for the nominal plant dynamics.
Initial parameter $\theta_{3} (0)$ is selected zero for all adaptive controllers.

\subsubsection{Selecting the design parameters for MRAC} \label{sec:rates}
Adaptation rate for a particular parameter $\theta_i$ in MRAC is chosen according to the empirical formula \cite{Yil:11} 
\begin{equation} \label{eq:rates}
\bar{\Gamma}_{ii}=\frac{|\theta_i^*|}{3 \tau_m (\bar{r})^2}
\end{equation}
where $\theta_i^*$ is the ideal value of the controller parameter, $\tau_m$ is the smallest time constant of the reference model and $\bar{r}$ is the maximum possible amplitude of the reference signal which is equal to the amplitude of the operating range. Since the ideal controller parameters, $\theta_i^*$, are assumed to be unknown, nominal values calculated using the matching conditions are used instead. 

Adaptation rates obtained from (\ref{eq:rates}) are calculated for the worst case scenario which is usually valid at the beginning of the operation when the tracking error and the system states are of the same order of magnitude with the reference signal. Moreover, the calculation (\ref{eq:rates}) requires the estimation of the ideal control parameters. Due to these approximations, we include a fine-tuning matrix $W$ for MRAC to fine-tune the adaptation rates as $\bar{\Gamma}_W=\bar{\Gamma}W$ where
\begin{equation} \label{eq:finetune}
W = \begin{bmatrix}
p_1 \ \ \ 0 \ \ \ 0 \\ 0 \ \ \ p_2 \ \ \ 0 \\ 0 \ \ \ 0 \ \ \ p_3
\end{bmatrix}
\end{equation}
and adjustable constants $p_{1,2,3}$ are used for the fine-tuning process. It was experienced during simulations that selection of $p_1=p_2=1$ and $p_3 > 1$ provide faster and more robust system response.

\subsubsection{Selecting the design parameters for CRM and DR-CRM adaptive controllers} \label{sec:ell}
CRM gain, $\ell$, helps suppress the oscillations in the case of high adaptation rates. However, a numerically large $\ell$ can cause so-called ``peak phenomena'' in non-zero initial tracking error \cite{Gibson:phd}. Therefore, a procedure to determine the optimum value of the CRM gain and adaptation rates is needed to reduce the time and effort spent for the controller tuning. Below steps, inspired from  \cite{Gibson:12}, are followed separately for CRM and DR-CRM adaptive controllers: 
\begin{enumerate}
\item Find the adaptation rates using (\ref{eq:rates}) and define the adaptation rate vector, $\bar{\gamma}$. 
\item Find the norm of the adaptation rate vector, $||\bar{\gamma}||$.
\item Define the CRM gain, $\ell$, equal to $||\bar{\gamma}||$.
\item Increase $\ell$ and $\bar{\gamma}$ together by keeping $\ell=||\bar{\gamma}||$ until a desired tracking performance is obtained in numerical simulations. 
\end{enumerate}

The same adaptation rate, $\gamma_{\lambda}$, which is determined through simulations, is chosen for each $\lambda_i$ since they have same order of magnitude.

\subsection{Step by step controller design procedure} \label{sec:procedure}
A clear, step by step procedure is provided below to facilitate the adaptive controller design.
\begin{steps}
	\item Determine the reference model dynamics by choosing appropriate ($A_{m},b_m,h_{m}$) for the performance specifications of the closed loop system. ($A_{m},b_m,h_{m}$) should form a strictly positive real (SPR) transfer function for MRAC.
	\item Signals $\omega_{1,2}(t)$ in (\ref{eq:MRAC_ss}) and (\ref{eq:plant_ss}) are generated by choosing ($F,g$) and ($\Lambda,b_{\Lambda}$) so that they form controllable pairs. Since these signals operate like state observers, eigenvalues of the matrices $F$ and $\Lambda$ should be faster than the reference model dynamics. For a first order plant, these signals are not required.
	\item Set the initial conditions of the controller parameters using Section \ref{sec:init}.
	\item Find the adaptation rates and CRM gain, $\ell$, using Section \ref{sec:rates}-\ref{sec:ell}. CRM gain should be selected such that $W_e(s)$ in (\ref{eq:SPR}) is an SPR transfer function for both CRM and DR-CRM adaptive controllers. 
	\item Tune the parameters $\gamma_{\lambda}$ and $p_3$ in (\ref{eq:discretized_vector}) and (\ref{eq:finetune}) using numerical simulations and then fine-tune again during experiments.
	\item Integrate the projection algorithm provided in (\ref{eq:projection_adaptivelaws}) to the adaptation laws.
\end{steps}

All of the adaptive controllers considered in this paper requires minimal amounts of computational resources and memory. DR-CRM, having the highest number of terms in the control signal, for example, needs only 256 bytes of memory for data storage. It has 116 operations per cycle, which corresponds to around 2320 floating point operations per second (flops). See the Appendix \ref{sec:appA} for detailed memory requirement and computational load calculations.

\section{Simulations} \label{sec:sim}
Simulation results employing the full nonlinear model, which is developed in Section \ref{sec:model}, are presented in this section. All the parameters defining the controller (initial conditions of the controller parameters, adaptation rates, CRM gain, ($F,g$) and ($\Lambda,b_{\Lambda}$) pairs) are obtained using Section \ref{sec:procedure}. Reference model is chosen to satisfy the performance specifications listed in Table \ref{tab:specs}. The controller gains of the PI controller is designed so that the closed loop system provide a similar response with the reference model, as explained in Section \ref{sec:PI}.

\begin{table} [htp]
	\centering
	\caption{Specifications of chosen reference model}
	\label{tab:specs}
	\begin{tabular}{|l|l|l|l|}
		\hline
		Steady state error & Rise time & Settling time (5 \%) & Maximum overshoot \\ \hline
		0 \% & 0.6 sec & 1.5 sec & 10 \% \\ \hline
	\end{tabular}
\end{table}

Firstly, performance of MRAC is compared with the PI controller. Then, MRAC, CRM adaptive controller and DR-CRM adaptive controller are comparatively evaluated, where demanding tracking tasks are utilized to reveal the performance differences between these controllers. Numerical simulations are carried out using Matlab \textregistered \ with a sampling interval of $50$ ms.

\subsection{MRAC vs. PI controller}

\begin{figure} [htp]
	\centering
	\includegraphics[trim = 0mm 99mm 0mm 99mm, clip, width=1\textwidth]{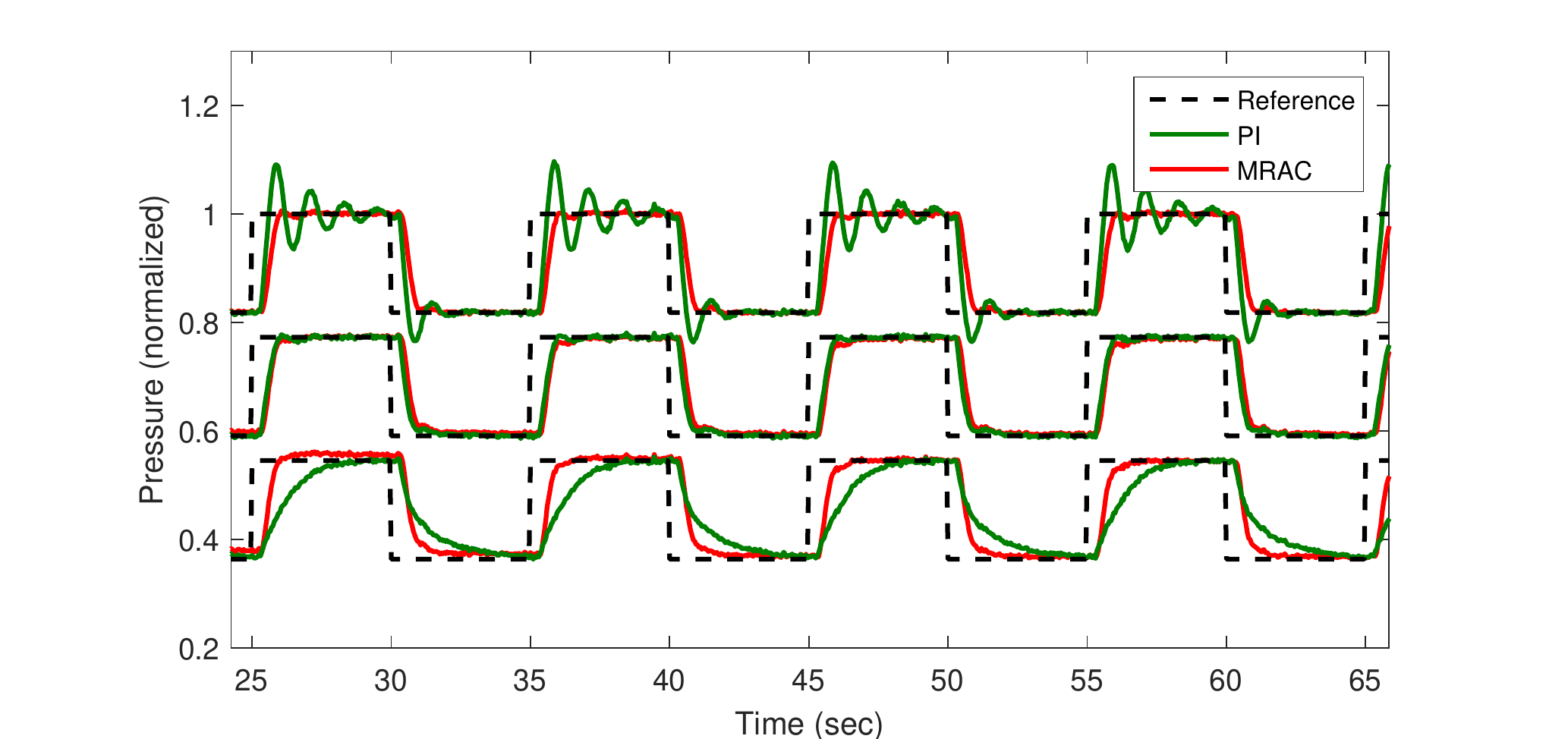}
	\caption{Tracking curves for the PI controller and MRAC at 3 different operating points. }
	\label{fig:Y1}
\end{figure}
In Fig. \ref{fig:Y1}, simulation results are given demonstrating the performances of MRAC and the PI controller. It is noted that although the PI controller and MRAC show very similar performances around the normalized nominal pressure, between 0.6 and 0.8, the closed loop system with the PI controller presents an oscillatory response for higher pressure operating conditions and a slow response for lower pressures. On the other hand, MRAC can adapt itself to changing operating conditions and provides a more consistent performance across operating points.

\subsection{Comparative evaluation of adaptive controllers}

\begin{figure}[htp]
	\centering
	\includegraphics[trim = 0mm 99mm 0mm 99mm, clip, width=1\textwidth]{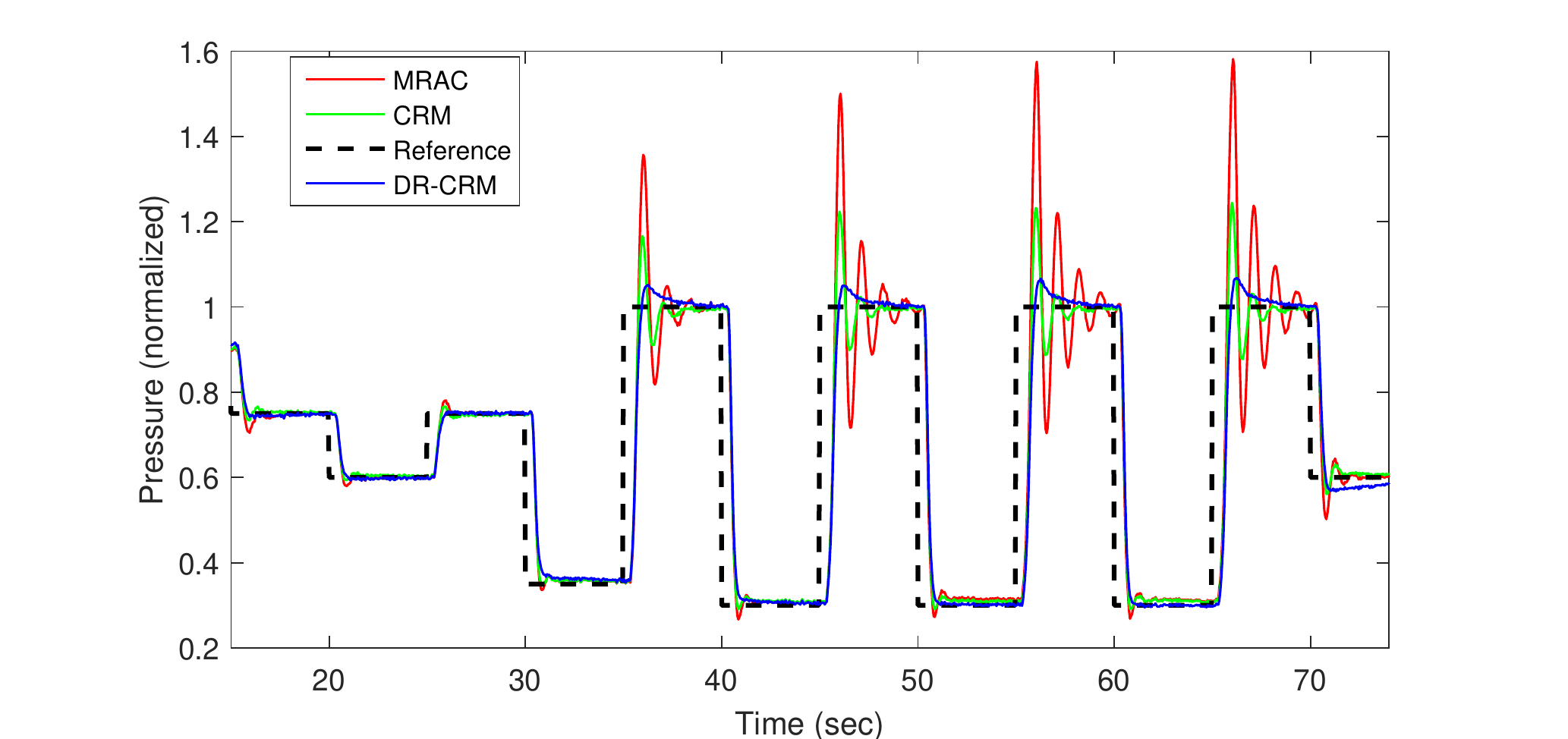}
	\caption{Reference tracking of MRAC, CRM adaptive control and DR-CRM adaptive control in simulations.}
	\label{fig:Y2}
\end{figure}

\begin{figure}[htp]
	\centering
	\includegraphics[trim = 0mm 99mm 0mm 99mm, clip, width=1\textwidth]{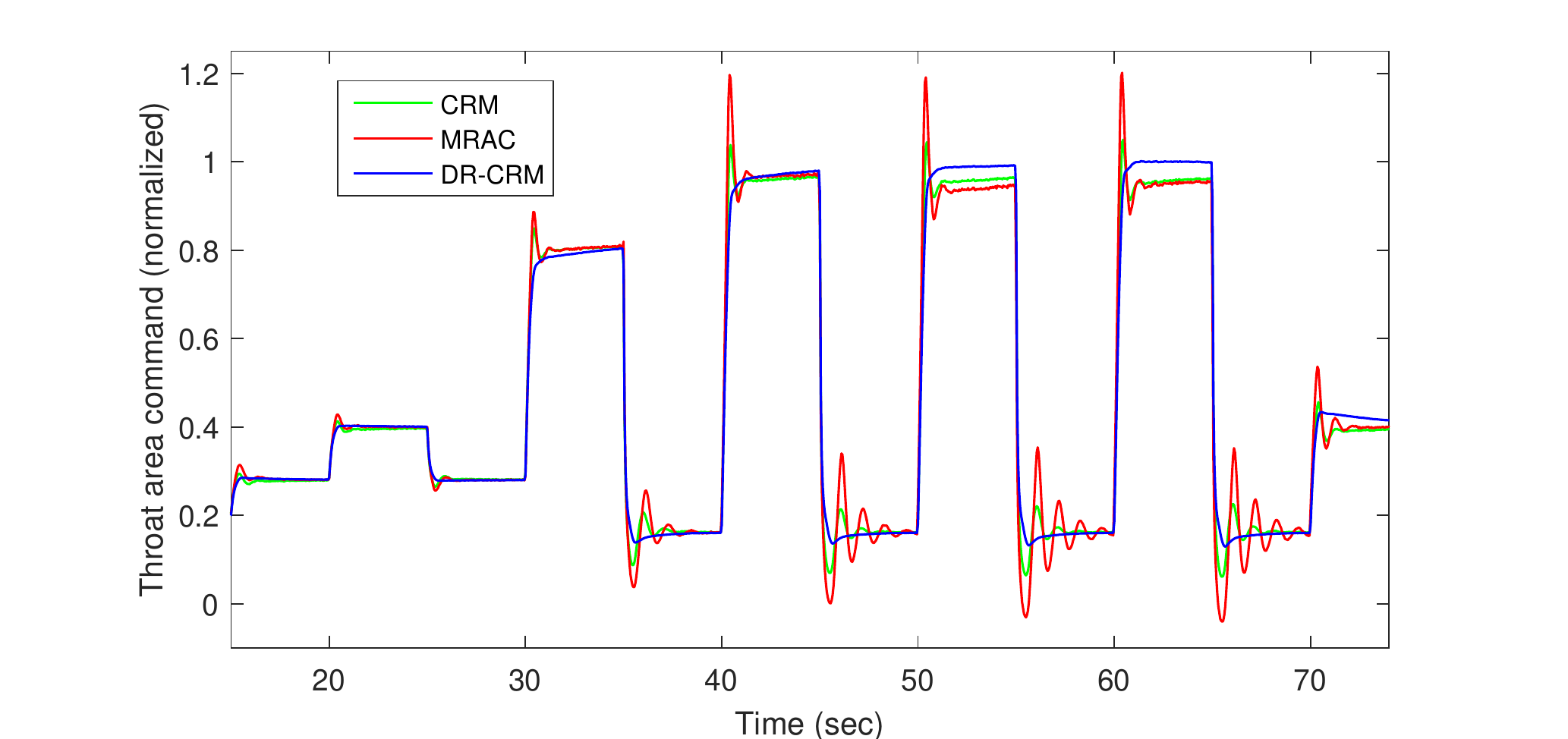}
	\caption{Evolution of control inputs of MRAC, CRM adaptive control and DR-CRM adaptive control in simulations.}
	\label{fig:U2}
\end{figure}

\begin{figure}[htp]
	\centering
	\includegraphics[trim = 0mm 99mm 0mm 99mm, clip, width=1\textwidth]{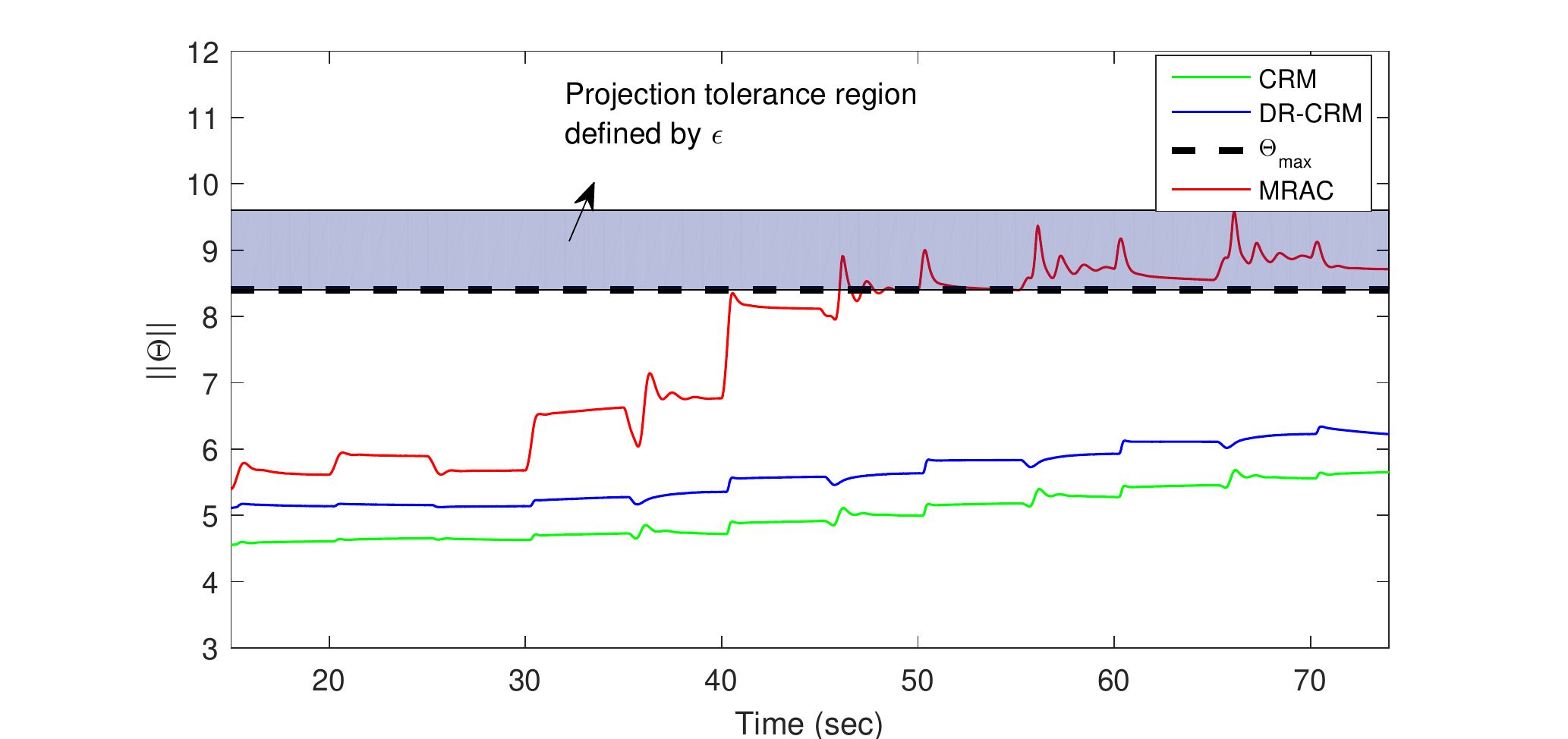}
	\caption{Controller parameters of MRAC, CRM adaptive control and DR-CRM adaptive control in simulations with the projection boundary}
	\label{fig:K2}
\end{figure}

The promise of CRM and delay compensation modifications for the conventional MRAC controller is that they provide higher performance without causing excessive oscillations. To demonstrate the difference that these modifications can make, more demanding reference trajectories, compared to those of the case presented in Fig. \ref{fig:Y1}, are used for the reference tracking tests. Also, time constant of the reference model, $\tau_m$, is halved to obtain a fast response. Adaptation rates of MRAC are updated using (\ref{eq:rates}) and a new $p_3$, employed in (\ref{eq:finetune}), is obtained empirically. The results of these simulations are given in Fig. \ref{fig:Y2}-\ref{fig:K2}.

MRAC, CRM adaptive controller and DR-CRM adaptive controller provide similar performances for small variations in the pressure demand but MRAC response becomes oscillatory once the demanded variation is increased by 3 times. Although CRM adaptive controller provides a considerably more damped response compared to MRAC, the best response is obtained for the case where DR-CRM adaptive controller is employed. Figure \ref{fig:U2} shows that DR-CRM controller provides the smoothest control input. In Fig. \ref{fig:K2}, adaptive control parameters are shown where MRAC parameters hit the projection boundary and stays within the projection tolerance limits. 

\section{Experiments} \label{sec:experiment}
\begin{figure} [htp]
	\centering
	\includegraphics[trim = 40mm 15mm 40mm 30mm, clip, width=1\textwidth]{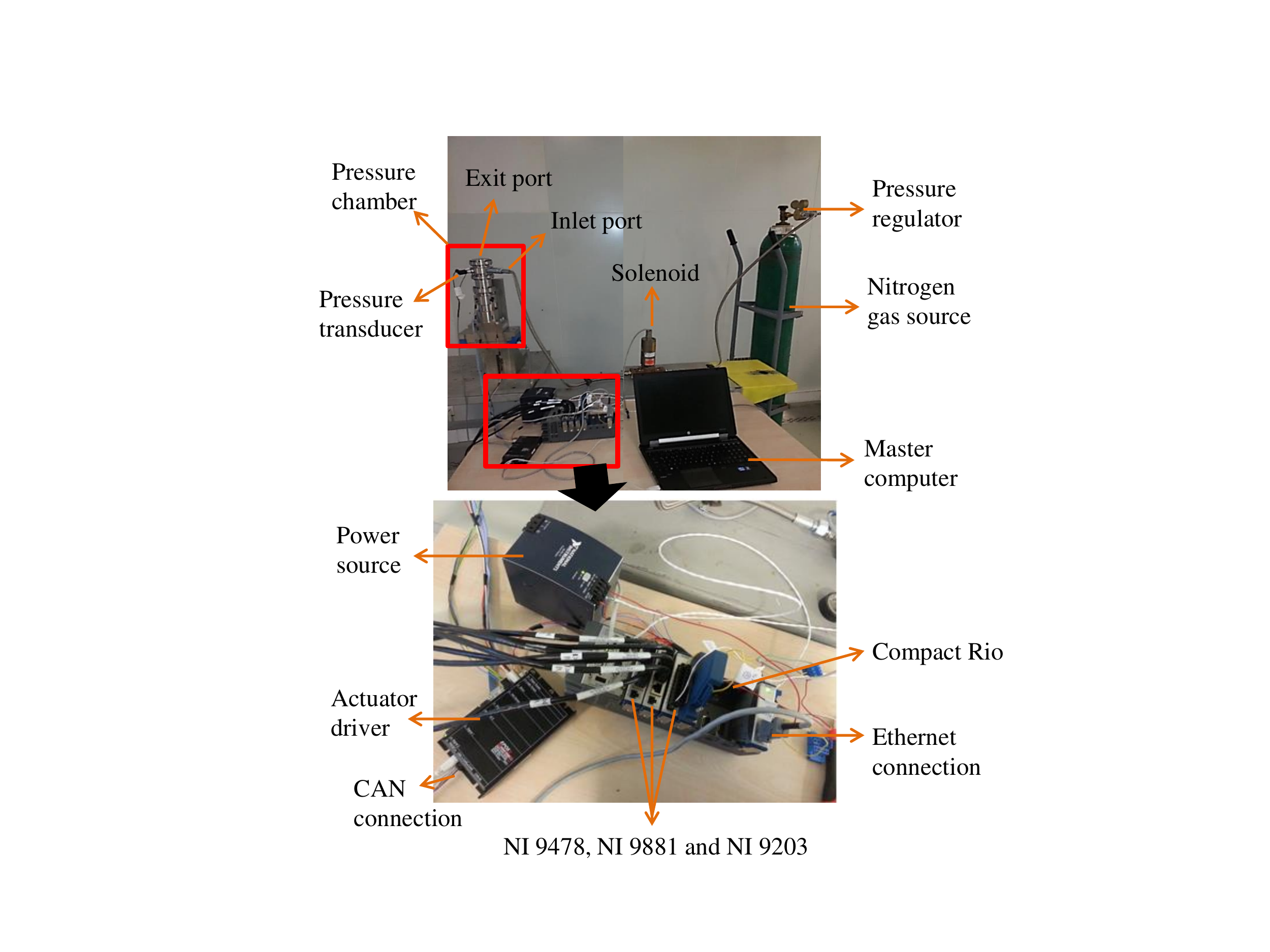}
	\caption{Cold Air Test Setup.}
	\label{fig:ExpSetup2}
\end{figure}
\begin{figure} [htp]
	\centering
	\includegraphics[trim = 20mm 70mm 0mm 20mm, clip, width=1\textwidth]{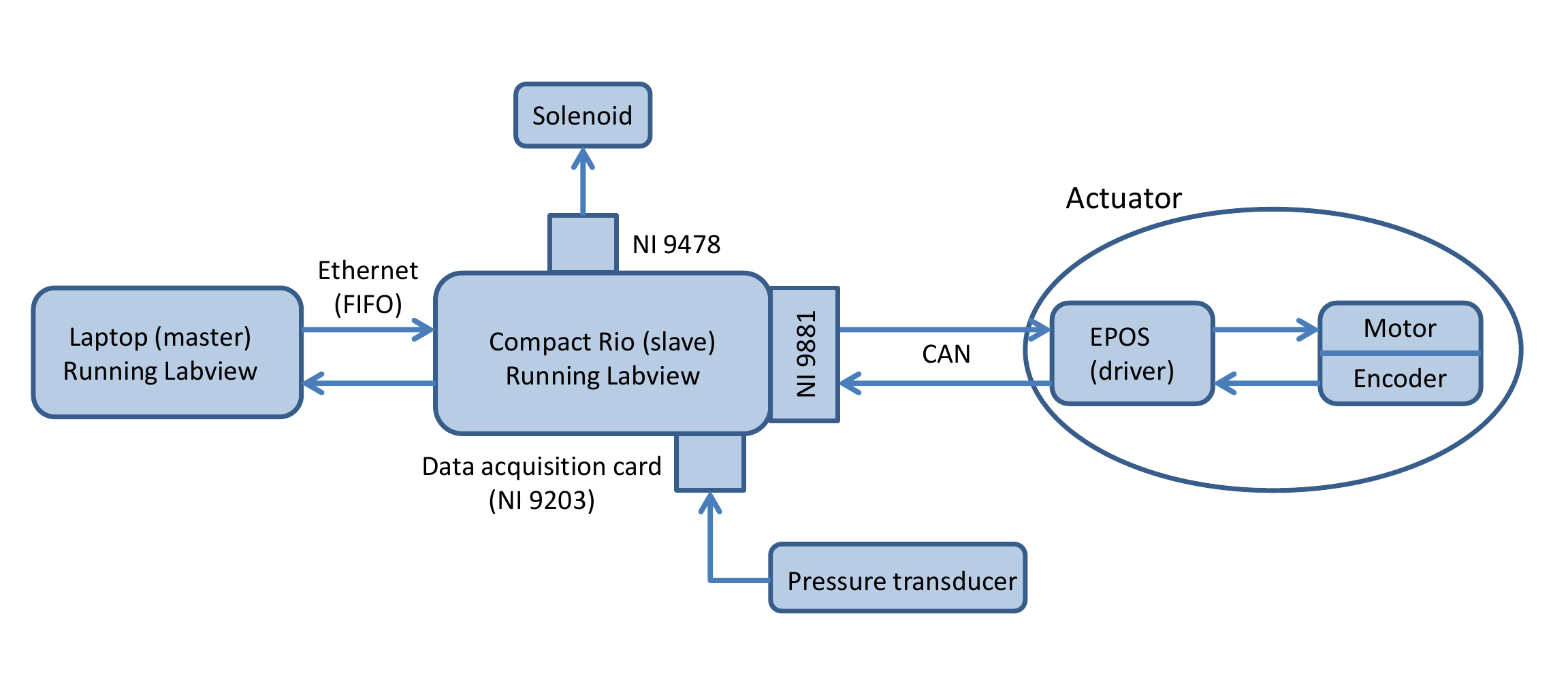}
	\caption{Schematic of the CATS hardware and data communication.}
	\label{fig:ExpSetup1}
\end{figure}
The experimental results are obtained using a cold air test setup (CATS), which is designed and manufactured by Roketsan Missiles Inc. (See Fig. \ref{fig:CATP} for a schematic of the overall system and Fig. \ref{fig:ExpSetup2} for the real experimental system). The pressure chamber in CATS has two ports, which are called the inlet port and the exit port. The inlet port is connected to a nitrogen gas source of 230 bars through a pressure regulator. The pressure regulator ensures safe test conditions by adjusting the inflow pressure. There is a solenoid between the pressure chamber and the pressure regulator to stop the flow in case of emergencies. Outlet port of the pressure chamber has a shape of a nozzle whose effective throat area is continuously altered during the operation by the actuator and valve mechanism. An EC-max 30, 60 Watt, 24 Volt brushless direct current motor (Maxon Motor Company \textregistered) with a EPOS2 70/10 driver is used as the actuator. Output shaft of the motor is connected to a gear box and a spindle, respectively. Other side of the spindle is connected to a conical pintle which is located such that its linear position determines the effective throat area at the outlet of the pressure chamber (see Fig. \ref{fig:valve}). There is a pressure transducer inside the pressure chamber, which provides real time pressure data to the controller. A slave Compact Rio computer (National Instrument \textregistered), running LabVIEW software, is used to collect data from the pressure transducer, run the pressure controller cycles to calculate the necessary effective throat area and send this data to the actuator driver which is responsible for controlling the actuator position. The communication between the slave Compact Rio computer and the actuator driver card is through NI 9881 card using a CAN bus protocol. Pressure transducer is connected to the data acquisition card (NI 9203) that is also connected to the Compact Rio. All the algorithms for data acquisition, pressure controller calculations, data sending and corresponding communication phases are prepared in the master computer prior to the experiments using LabVIEW and the code is embedded to the slave Compact Rio computer through ethernet connection. Communication between the master and slave computers is realized with FIFO: First-In-First-Output methodology. Master computer monitors the experiment in real time and is able to intervene in the experimental process in case of a safety hazard. The solenoid located in between the pressure chamber and the pressure regulator is controlled by the master computer through the digital card NI 9478. A detailed schematic of the hardware and data communication are given in Fig. \ref{fig:ExpSetup1}.

The same scenarios used for the simulations are employed for the experimental tests. First, better performance of MRAC over the PI controller is demonstrated by performing experiments at three different operation points with the same controller gains and parameters used in the simulations. Then, a comparative evaluation showing the advantage of DR-CRM adaptive controller over other adaptive controllers are presented. Finally, an experiment is conducted for a larger period of time to show the effectiveness of the projection algorithm. All the numbers in the figures are normalized. 

\subsection{MRAC vs. PI controller}
\begin{figure} [htp]
	\centering
	\includegraphics[trim = 0mm 99mm 0mm 99mm, clip, width=1\textwidth]{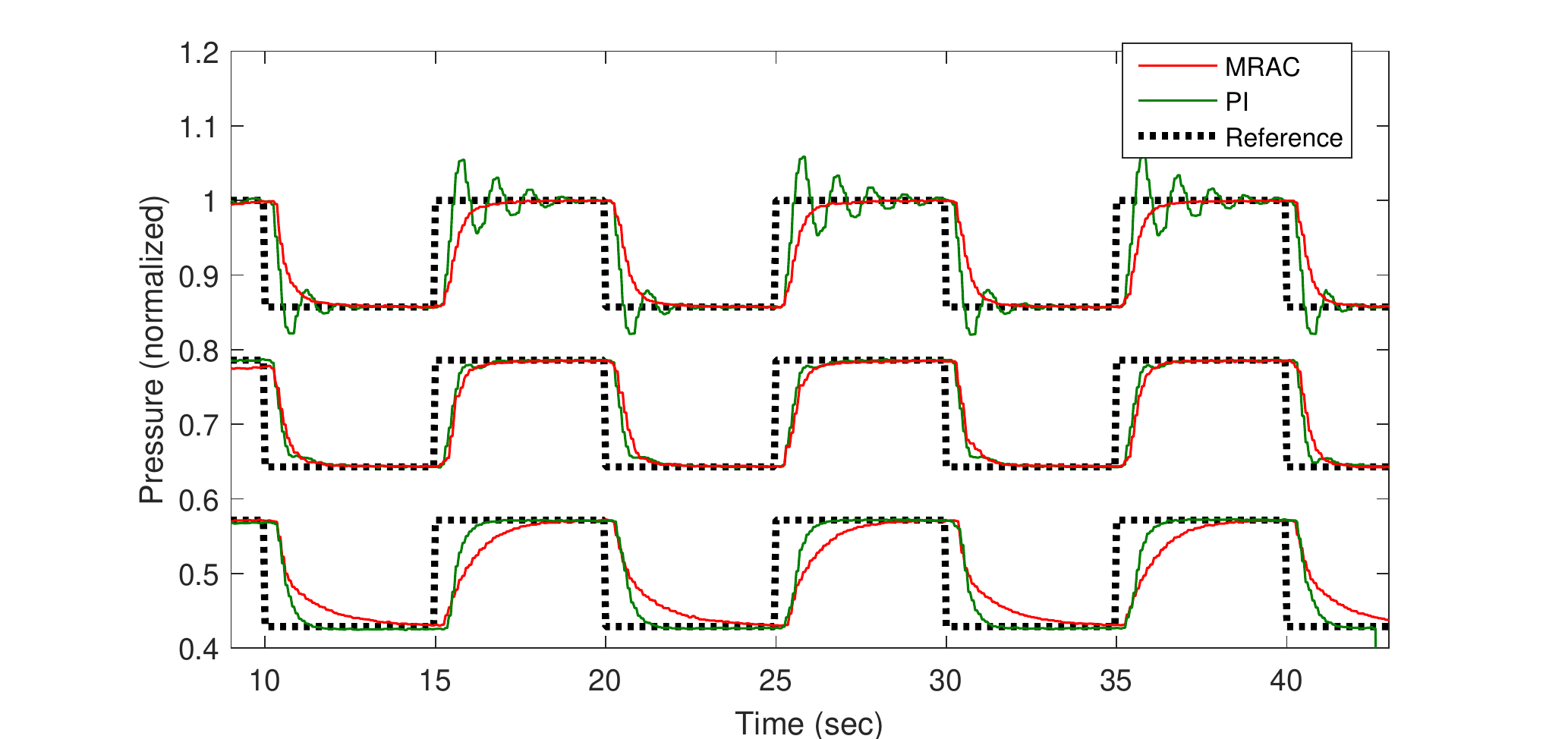}
	\caption{Test results of PI controller and MRAC for three different operating conditions. The adaptation rates used for these experiments are the same as the ones used in simulations.}
	\label{fig:Ytest1}
\end{figure}
Test results are given in Fig. \ref{fig:Ytest1}. PI controller shows acceptable performance at the nominal operating point, the linearized model of which was used for the controller design. However, as the operating point deviates from the nominal design conditions, advantage of the adaptive controller, which provides consistent transient performance at different operating conditions, is observed. 

\subsection{Comparative evaluation of adaptive controllers}
To demonstrate the performance differences between the adaptive controllers, a more challenging reference signal, compared to that of the previous subsection, is used and time constant of the reference model, $\tau_m$, is halved. Same controller design parameters are used as in the simulations. 

\begin{figure}[htp]
	\centering
	\includegraphics[trim = 0mm 99mm 0mm 99mm, clip, width=1\textwidth]{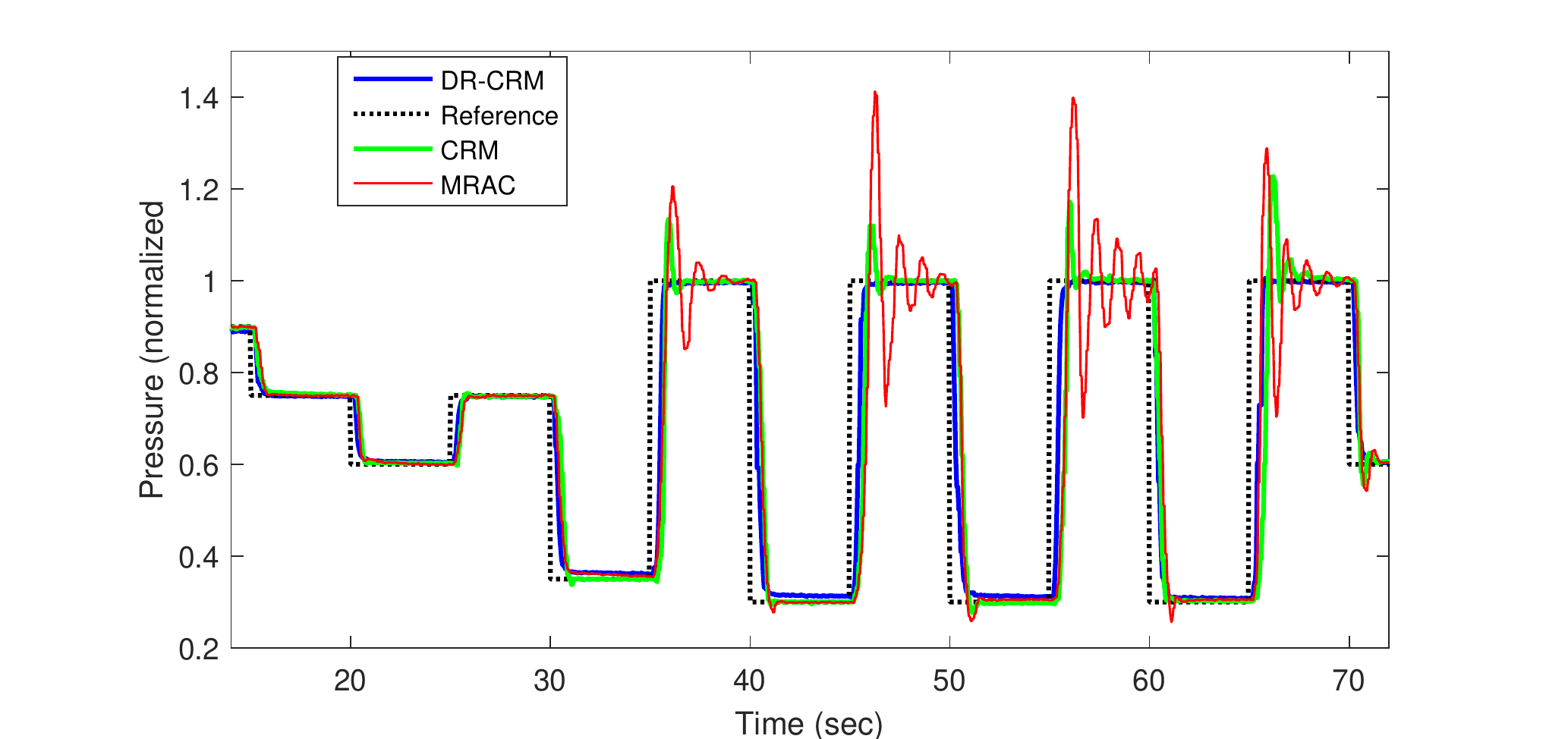}
	\caption{Reference tracking of MRAC, CRM adaptive control and DR-CRM adaptive control in experiments.}
	\label{fig:Ytest3}
\end{figure}

\begin{figure}[htp]
	\centering
	\includegraphics[trim = 0mm 99mm 0mm 99mm, clip, width=1\textwidth]{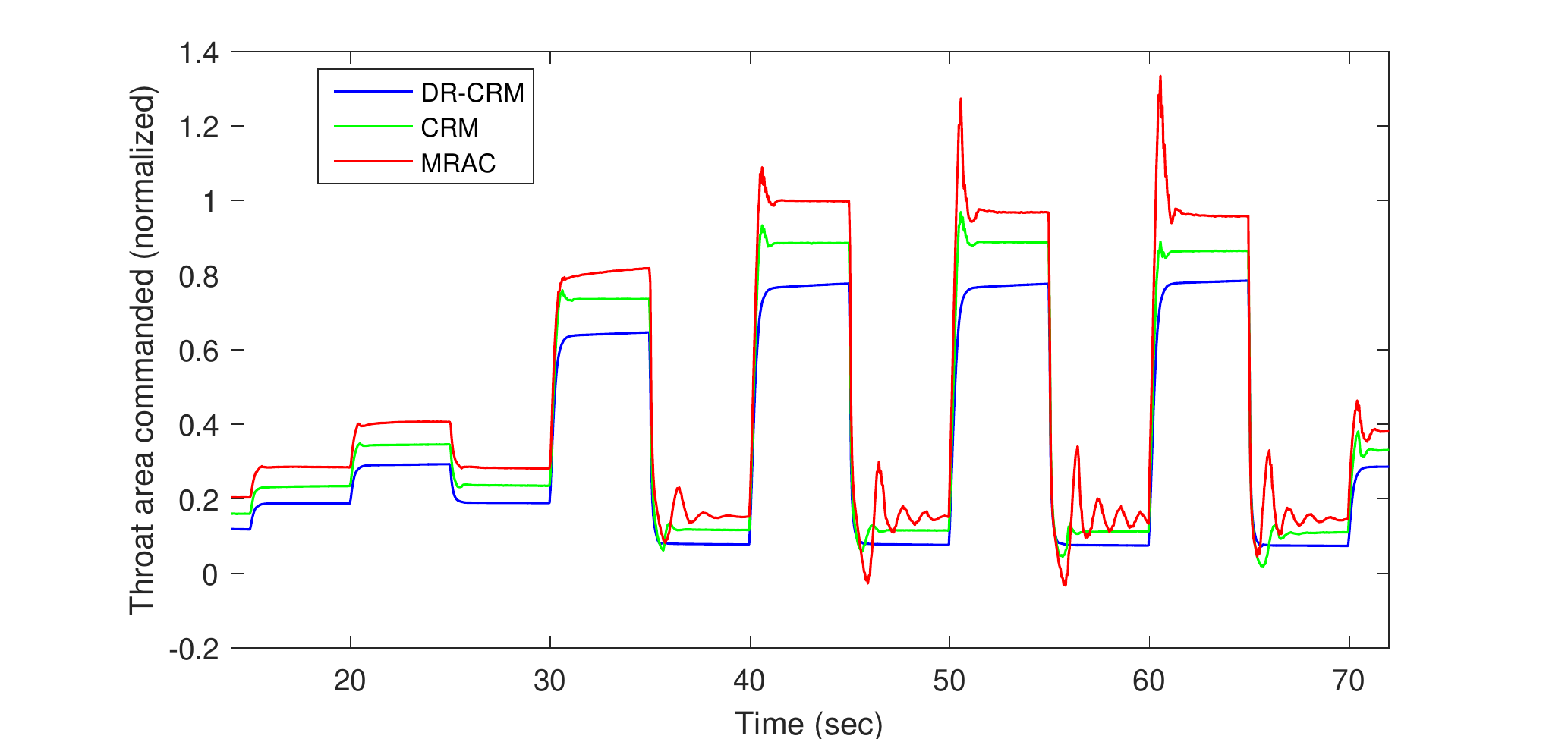}
	\caption{Evolution of control inputs of MRAC, CRM adaptive control and DR-CRM adaptive control in experiments.}
	\label{fig:Utest3}
\end{figure}

\begin{figure}[htp]
	\centering
	\includegraphics[trim = 0mm 99mm 0mm 99mm, clip, width=1\textwidth]{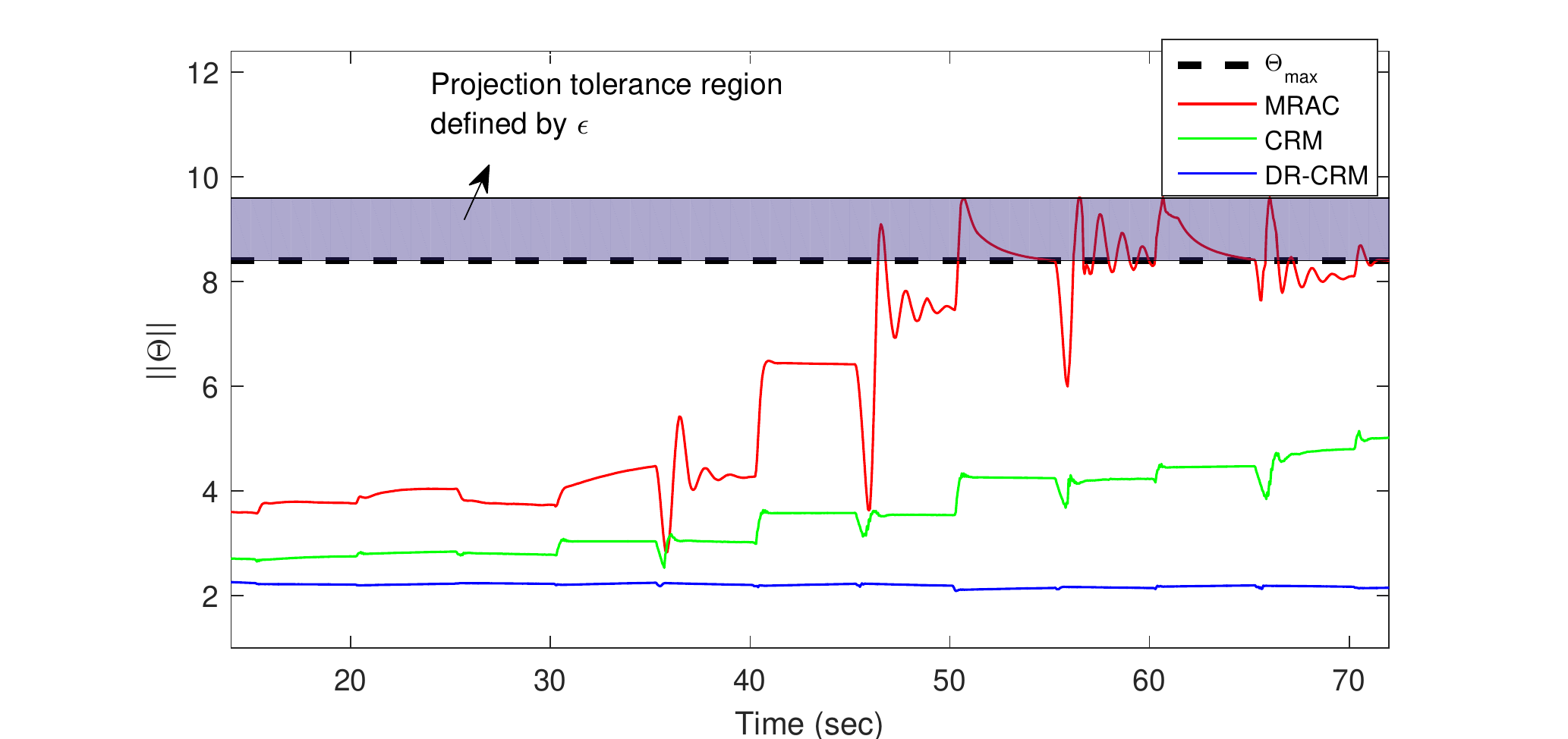}
	\caption{Controller parameters of MRAC, CRM adaptive control and DR-CRM adaptive control in experiments and projection boundary}
	\label{fig:Ktest3}
\end{figure}

Experiment results are given in Fig. \ref{fig:Ytest3}-\ref{fig:Ktest3}, exhibiting similar trends with the simulations with CRM and DR-CRM adaptive controller having slightly more damped responses. DR-CRM adaptive controller is able handle the demanding operation conditions in the experiments and provides a reasonable performance. CRM adaptive controller, damping most of the oscillations as intended, results in undesired high amplitude overshoots. MRAC's response is similar to that of the other two adaptive controllers for small pressure deviation demands but becomes oscillatory for larger deviations in the reference signal. DR-CRM adaptive controller has the smoothest controller input, exhibited in Fig. \ref{fig:Utest3}. Fig. \ref{fig:Ktest3} shows that the controller parameters of MRAC hit the projection boundary, but then prevented to grow further.

\begin{figure} [htp]
	\centering
	\includegraphics[trim = 0mm 99mm 0mm 99mm, clip, width=1\textwidth]{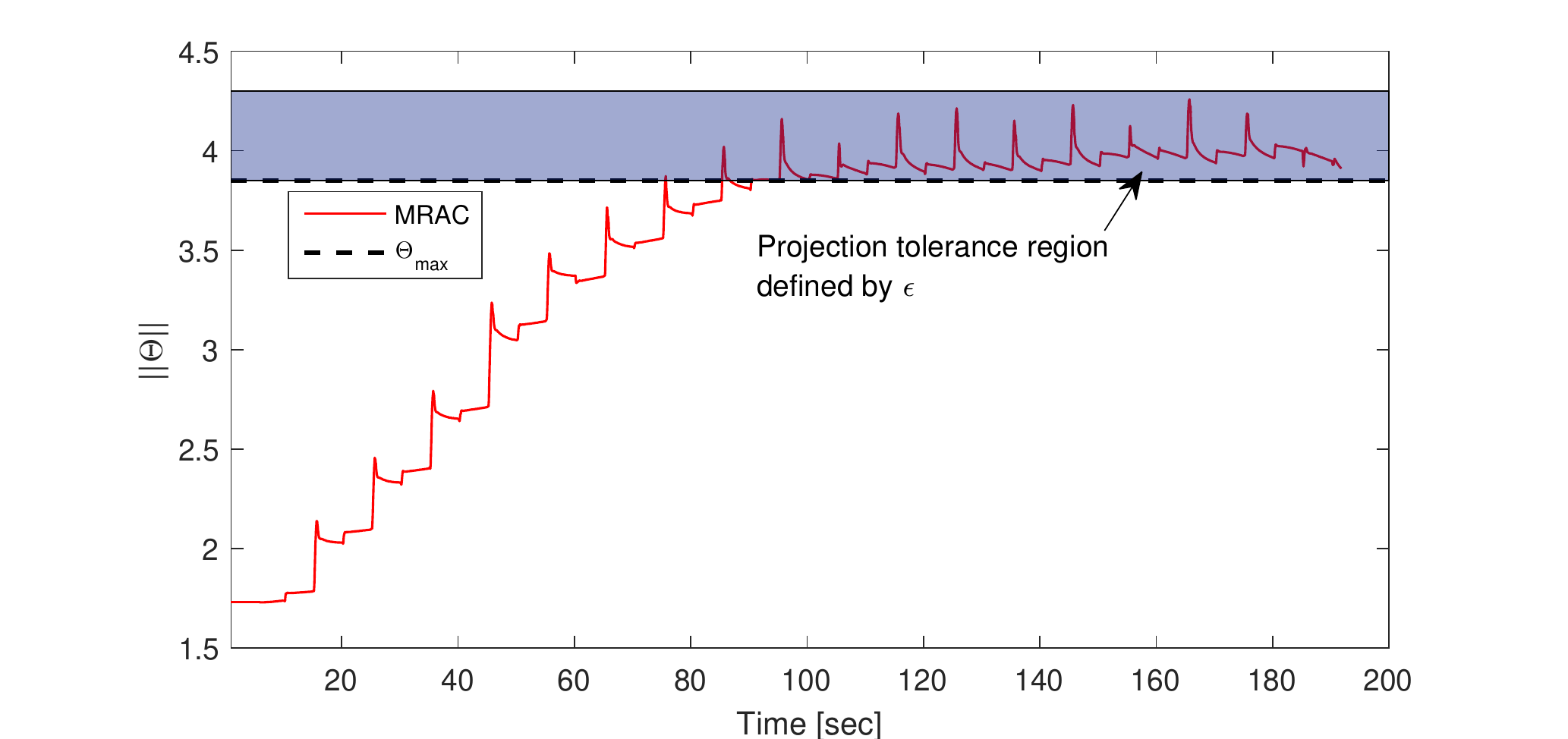}
	\caption{Controller parameters in the long-term test}
	\label{fig:Ktest4}
\end{figure}

\begin{figure} [htp]
	\centering
	\includegraphics[trim = 0mm 99mm 0mm 99mm, clip, width=1\textwidth]{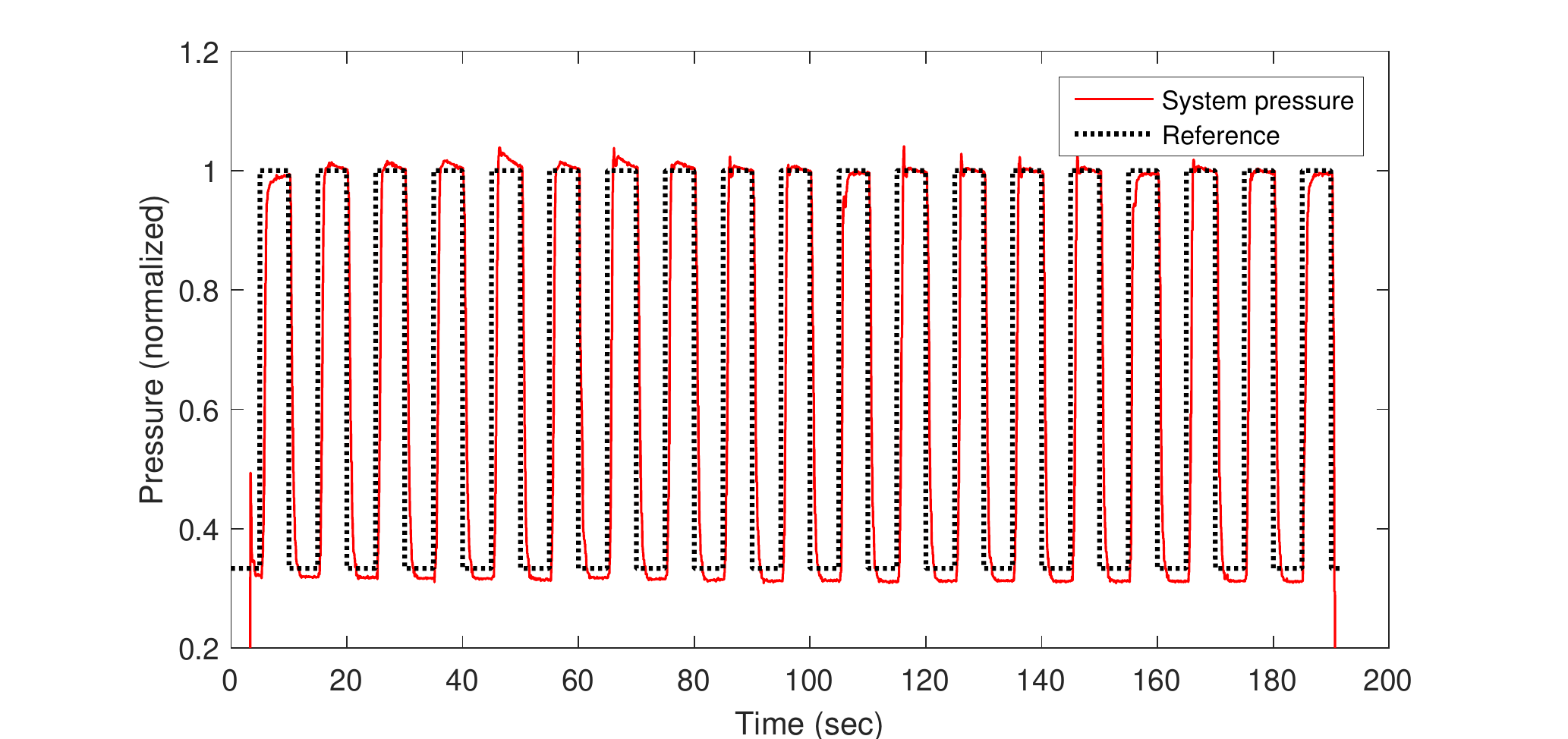}		
	\caption{Pressure tracking in the long-term test}
	\label{fig:Ytest4}
\end{figure}

Effect of the projection algorithm can be observed more clearly in the experiments that are conducted for longer times. Fig. \ref{fig:Ktest4} presents the evolution of the norm of the MRAC controller parameters in a longer test whose tracking curve is depicted in Fig. \ref{fig:Ytest4}. Controller parameters tend to increase due to non-ideal situations such as unmodeled dynamics, disturbances and noise, but the projection algorithm keeps them within a predefined bound. 


\section{Summary and Future Work} \label{sec:conc}
The pressure control of gas generators in ducted rockets are addressed in this paper. To solve this control problem, which includes time delays, uncertainties and nonlinear dynamics, a delay resistant closed loop reference model (DR-CRM) adaptive controller is proposed. The controller merges the benefits of two different approaches: The first approach is the adaptive posicast controller (APC) which compensates the time delay by making use of positively forecasted output of the plant. The second approach is the closed-loop reference model (CRM) modification, which damps the high frequency oscillations due to high adaptive learning rates by altering the reference model structure with tracking error feedback. 

DR-CRM adaptive controller is tested using a cold air test setup (CATS) which is utilized as a test bed for throttleable ducted rocket development. A detailed nonlinear mathematical model of the CATS is presented. Model reference adaptive control (MRAC) and CRM adaptive control along with a proportional integral (PI) controller are also designed for a comparative evaluation with the proposed controller. Implementation enhancements are presented to ensure the robustness of the controllers against non-ideal experimental conditions. A step by step controller design procedure is also provided.

Performances of all controllers are evaluated in simulations and experiments, and similar conclusions were drawn for both cases. MRAC clearly demonstrates its advantage over constant gain PI controller. More demanding conditions reveal the advantage of DR-CRM adaptive controller over CRM adaptive controller and MRAC. Effect of the projection algorithm to ensure the boundedness of the controller parameters are also shown in the experiments.

The results of this research showed that combining different control approaches is effective in addressing complex control problems such as gas generator pressure control. As a future work, the authors plan to continue this research with experimental studies on real gas generators and continue to improve/modify the proposed approach according to possibly new implementation requirements. These requirements may include time-varying or uncertain time delays, the need for discrete domain design due to non-ideal sampling or different kind of disturbances that require more sophisticated disturbance rejection methods. 

\newpage

\appendix* 
\section{Memory requirement and computational load of the implementation of the DR-CRM adaptive controller} \label{sec:appA}
Input time delay is observed to be 300 ms on the average, whereas sampling interval of the controller cycle is chosen as 50 ms, which adds 6 controller parameters, $\lambda_i$, and 6 states, $u(t-mdt)$, to the controller structure (see (\ref{eq:discretized_vector})). There are 9 states overall, 9 controller parameters, 9 multiplication results, which adds up to 1 controller signal. We need 9 terms to the define adaptation laws, 4 terms to define reference model ($a_m, b_m,\ell$ and $y_m$), 1 tracking error term and 9 adaptation rate terms. In addition, we need 13 terms for the projection algorithm ($\Theta_{max},\epsilon,||\Theta||,f$ and $\nabla f$). Overall, we have 64 float variables that needs 256 bytes of memory space. 

9 multiplication and 8 summation operations are needed to define the controller signal. There exist 18 multiplications in the adaptive law calculations and 9 summation operations are required to update the controller parameters. In addition, 4 summation and 3 multiplication operations are needed to be employed to form the reference model output and the tracking error. Furthermore, in the projection algorithm, there are 2 comparison and 1 logical operations, 18 summation and 43 multiplication operations along with a square root operator. In total, 116 floating point operations are conducted per controller cycle that runs with a sampling rate of 50 ms, which results in 2320 floating point operations per second (FLOPS).

\bibliographystyle{aiaa}
\bibliography{references} 

\begin{thebibliography}{10}
\newcommand{\enquote}[1]{``#1''}
\expandafter\ifx\csname urlstyle\endcsname\relax
  \providecommand{\doi}[1]{doi:\discretionary{}{}{}#1}\else
  \providecommand{\doi}{doi:\discretionary{}{}{}\begingroup
  \urlstyle{rm}\Url}\fi

\bibitem{Besser:12}
Besser, H.-L. and Kurth, G., \enquote{METEOR - European Air Dominance Missile
  Powered by High Energy Throttleable Ducted Rocket,} in \enquote{Proc.
  RTO-MP-AVT-208,} , 2012, pp. 1--17.

\bibitem{Goldman}
Goldman, C. and Gany, A., \enquote{Thrust modulation of ram-rockets by a vortex
  valve,} in \enquote{AIAA, ASME, SAE, and ASEE, Joint Propulsion Conference
  and Exhibit,} Lake Buena Vista, FL, 1996.

\bibitem{Miller:81}
Miller, W., McClendon, S., and Burkes, W., \enquote{Design Approaches for
  Variable Flow Ducted Rockets,} in \enquote{Proc. AIAA/SAE/ASME 17th Joint
  Propulsion Conference,} Colorado Springs, Colorado, 1981.

\bibitem{Davis:03}
Davis, C.~A. and Gerards, A.~B., \enquote{VARIABLE THRUST SOLID PROPULSION
  CONTROL USING LABVIEW,} in \enquote{Proc. AIAA/ASME/SAE/ASEE 39th Joint
  Propulsion Conference and Exhibit,} Huntsville, Alabama, AIAA 2003-5241,
  2003.

\bibitem{Chang:11}
Chang, J., Li, B., Bao, W., Niu, W., and Yu, D., \enquote{Thrust control system
  design of ducted rockets,} \emph{Acta Astronautica}, Vol.~69, No.~1, 2011,
  pp. 86--95.

\bibitem{Burroughs}
Burroughs, S., \enquote{Status of Army Pintle Technology for Controllable
  Thrust Propulsion,} in \enquote{37th AIAA/ASME/SAE/ASEE Joint Propulsion
  Conference and Exhibit,} Salt Lake City, Utah, 2001.

\bibitem{Ostrander:97}
Ostrander, M. and Thomas, M., \enquote{Air Turbo-Rocket solid propellant
  development and testing,} in \enquote{Proc. AIAA/ASME/SAE/ASEE 33th Joint
  Propulsion Conference and Exhibit,} , 1997.

\bibitem{Bao:10}
Bao, W., Li, B., Chang, J., Niu, W., and Yu, D., \enquote{Switching control of
  thrust regulation and inlet buzz protection for ducted rocket,} \emph{Acta
  Astronautica}, Vol.~67, No.~7, 2010, pp. 764--773.

\bibitem{Bao:11}
Bao, W., Qi, Y., and Chang, J., \enquote{Multi-objective regulating and
  protecting control for ducted rocket using a bumpless transfer scheme,}
  \emph{Proceedings of the IMechE, Part G:Journal of Aerospace Engineering},
  pp. 311--325.

\bibitem{Chang:14}
Qi, Y., Bao, W., Zhao, J., and Chang, J., \enquote{Coordinated control for
  regulation/protection mode-switching of ducted rockets,} \emph{Acta
  Astronautica}, Vol.~98, 2014, pp. 138--146.

\bibitem{Thomaier:87}
Thomaier, D., \enquote{Speed control of a missile with throttleable ducted
  rocket propulsion,} in \enquote{Proc. Advances in Air-Launched Weapon
  Guidance and Control 15 p (SEE N88-19553 12-15),} , 1987.

\bibitem{Sreeriatha:99}
Sreeriatha, A.~G. and Bhardwaj, N., \enquote{Mach Number Control-ler for a
  Flight Vehicle with Ramjet Propulsion,} in \enquote{Proc. AIAA/ASME/SAE/ASEE
  35th Joint Propulsion Conference and Exhibit,} Los Angeles, CA, AIAA 99-294,
  1999.

\bibitem{Pinto:11}
Pinto, P. and Kurth, G., \enquote{Robust Propulsion Control in all Flight
  Stages of a Throtteable Ducted Rocket,} in \enquote{Proc. AIAA/ASME/ASEE
  Joint Propulsion Conference \& Exhibit,} San Diego, California,
  AIAA-2011-5611, 2011, pp. 1--12.

\bibitem{Bao:2010}
Bao, W., Niu, W., J.T., C., Cui, T., and Yu, D., \enquote{Control system design
  and experiment of needle-type gas regulating system for ducted rocket,}
  \emph{Proceedings of the Institution of Mechanical Engineers, Part G, Journal
  of Aerospace Engineering}, Vol. 224, No.~5, 2010, pp. 563--573.

\bibitem{Niu:10}
Niu, W.~Y., Bao, W., Chang, J., Cui, T., , and Yu, D.~R., \enquote{Control
  system design and experiment of needle-type gas regulating system for ducted
  rocket,} \emph{Journal of Aerospace Engineering}, Vol. 224, No.~5, 2010, pp.
  563--573.

\bibitem{Bauer:12}
Bauer, C., Hopfe, N., Caldas-Pinto, P., Davenne, F., and Kurth, G.,
  \enquote{Advanced Flight Performance Evaluation Methods of Supersonic Air-
  Breathing Propulsion System by a Highly Integrated Model Based Approach,} in
  \enquote{Proc. RTO-MP-AVT-208,} , 2012, pp. 1--14.

\bibitem{Ilchmann:02}
Ilchmann, A., Ryan, E.~P., and Sangwin, C.~J., \enquote{Tracking with
  prescribed transient behaviour,} \emph{ESAIM: Control, Optimisation and
  Calculus of Variations}, Vol.~7, 2002, pp. 471--493.

\bibitem{Yil:Auto}
Yildiz, Y., Annaswamy, A., Kolmanovsky, I., and Yanakiev, D., \enquote{Adaptive
  Posicast Controller for Time-delay Systems with Relative Degree $n^*\leq2$,}
  \emph{Automatica}, Vol.~46, No.~2, 2010, pp. 279--289.

\bibitem{Gibson:12}
T.Gibson, A.Annaswamy, and E.Lavretsky, \enquote{Adaptive systems with
  closed-loop reference models: Stability, robustness, and transient
  performance,} \emph{arXiv:1201.4897. Ithaca, NY: Cornell University Library.
  Retrieved from http://arxiv.org/abs/1201.4897}.

\bibitem{Gibson:12_1}
T.Gibson, A.Annaswamy, and E.Lavretsky, \enquote{Improved Transient Response in
  Adaptive Control Using Projection Algorithms and Closed Loop Reference
  Models,} in \enquote{AIAA Guidance, Navigation, and Control Conference,}
  Minneapolis, Minnesota, AIAA-2012-4775, 2012.

\bibitem{Gibson:12_3}
T.Gibson, A.Annaswamy, and E.Lavretsky, \enquote{On Adaptive Control with
  Closed-Loop Reference Models: Transients, Oscillations, and Peaking,}
  \emph{IEEE Access}, Vol.~1, 2013, pp. 709--717.

\bibitem{Smith}
Smith, O.~J., \enquote{A controller to overcome dead time,} \emph{ISA Journal},
  Vol.~6.

\bibitem{Manitius}
Manitius, A.~Z. and Olbrot, A.~W., \enquote{Finite spectrum assignement problem
  for systems with delays,} \emph{IEEE Transactions on Automatic Control},
  Vol.~24, No.~4.

\bibitem{Ichikawa}
Ichikawa, K., \enquote{Frequency-domain pole assignement and exact
  model-matching for delay systems,} \emph{International Journal of Control},
  Vol.~41, 1985, pp. 1015--1024.

\bibitem{Ortega}
Ortega, R. and Lozano, R., \enquote{Globally stable adaptive controller for
  systems with delay,} \emph{International Journal of Control}, Vol.~47, No.~1,
  1988, pp. 17--23.

\bibitem{Sil:03}
Niculescu, S.-I. and Annaswamy, A.~M., \enquote{An adaptive Smith-controller
  for time-delay systems with relative degree $n^*\leq2$,} \emph{Systems and
  Control Letters}, Vol.~49, 2003, pp. 347--358.

\bibitem{YilJ:idle}
Yildiz, Y., Annaswamy, A., Yanakiev, D., and Kolmanovsky, I., \enquote{Spark
  Ignition Engine Idle Speed Control: An Adaptive Control Approach,} \emph{IEEE
  Transactions On Control Systems Technology}, Vol.~19, No.~5, 2011, pp.
  990--1002.

\bibitem{YilJ:FAR}
Yildiz, Y., Annaswamy, A., Yanakiev, D., and Kolmanovsky, I., \enquote{Spark
  Ignition Engine Fuel-to-Air Ratio Control: An Adaptive Control Approach,}
  \emph{Control Engineering Practice}, Vol.~18, No.~12, 2010, pp. 1369--1378.

\bibitem{Yil:07}
Yildiz, Y., Annaswamy, A., Yanakiev, D., and Kolmanovsky, I., \enquote{Adaptive
  Idle Speed Control for Internal Combustion Engines,} in \enquote{Proc. Amer.
  Control Conf.}, New York City, 2007, pp. 3700--3705.

\bibitem{Yil:08}
Yildiz, Y., Annaswamy, A., Yanakiev, D., and Kolmanovsky, I.,
  \enquote{Automotive Powertrain Control Problems Involving Time Delay: An
  Adaptive Control Approach,} in \enquote{Proc. ASME Dynamic Systems and
  Control Conference,} Ann Arbor, Michigan, 2008.

\bibitem{Yil:ACC08}
Yildiz, Y., Annaswamy, A., Yanakiev, D., and Kolmanovsky, I., \enquote{Adaptive
  Air Fuel Ratio Control for Internal Combustion Engines,} in \enquote{Proc.
  Amer. Control Conf.}, Seattle, Washington, 2008, pp. 2058--2063.

\bibitem{Dydek:13}
Z.T.Dydek, A.Annaswamy, J.J.E.Slotine, and E.Lavretsky, \enquote{Composite
  adaptive posicast control for a class of LTI plants with known delay,}
  \emph{Automatica}, Vol.~49, No.~6, 2013, pp. 1914--1924.

\bibitem{Pietri:auto}
Bresch-Pietri, D. and Krstic, M., \enquote{Adaptive trajectory tracking despite
  unknown input delay and plant parameters,} \emph{Automatica}, Vol.~45, No.~9,
  2009, pp. 2074--2081.

\bibitem{Liberis}
Bekiaris-Liberis, N. and Krstic, M., \enquote{Delay-adaptive feedback for
  linear feedforward systems,} \emph{System and Control Letters}, Vol.~59,
  No.~5, 2010, pp. 277--283.

\bibitem{Krstic:book}
Krstic, M., \emph{Delay Compensation for Nonlinear, Adaptive, and PDE Systems},
  Birkhauser, Boston, 2009.

\bibitem{Gibson:12_2}
T.Gibson, A.Annaswamy, and E.Lavretsky, \enquote{Closed-Loop Reference Models
  for Output-Feedback Adaptive Systems,} in \enquote{European Control
  Conference (ECC),} IEEE, Zurich, 2013, pp. 365--370.

\bibitem{Gibson:12_4}
T.Gibson, A.Annaswamy, and E.Lavretsky, \enquote{Adaptive Systems with
  Closed–loop Reference Models, Part I: Transient Performance,} in
  \enquote{American Control Conference,} IEEE, Piscataway, NJ, 2013, pp.
  3376--3383.

\bibitem{Lavretsky:12}
Lavretsky, E., \enquote{Adaptive output feedback design using asymptotic
  properties of lqg/ltr controllers,} \emph{IEEE Trans. Automat. Contr.},
  Vol.~57, No.~6.

\bibitem{Stepanyan:10}
Stepanyan, V. and Krishnakumar, K., \enquote{Mrac revisited: guaranteed
  perforamance with reference model modification,} in \enquote{American Control
  Conference,} , 2010.

\bibitem{Stepanyan:11}
Stepanyan, V. and Krishnakumar, K., \enquote{M–mrac for nonlinear systems
  with bounded disturbances,} in \enquote{Conference on Decision and Control,}
  , 2011.

\bibitem{Yucelen:14}
Yucelen, T., Torre, G. D.~L., and Johnson, E.~N., \enquote{Improving transient
  performance of adaptive control architectures using frequency-limited system
  error dynamics,} \emph{International Journal of Control}, Vol.
  DOI:10.1080/00207179.2014.922702.

\bibitem{Pomet:92}
Pomet, J.-B. and Praly, L., \enquote{Adaptive Nonlinear Regulation: Estimation
  from the Lyapunov Equation,} \emph{IEEE TRANSACTIONS ON AUTOMATIC CONTROL},
  Vol.~37, No.~6, 1992, pp. 729--740.

\bibitem{Lee:13catp}
Lee, J.~H., Park, B.~H., and Yoon, W., \enquote{Parametric investigation of the
  pintle-perturbed conical nozzle flows,} \emph{Aerospace Science and
  Technology}, Vol.~26, No.~5, 2013, pp. 268--279.

\bibitem{Deng:15}
Deng, R., Setoguchi, T., and Kim, H.~D., \enquote{Computational Study on the
  Thrust Performance of a Supersonic Pintle Nozzle,} in \enquote{International
  Symposium of Turbulence and Shear Flow Phenomena,} Melbourne, Australia,
  2015.

\bibitem{Heo:15}
Heo, J., Jeong, K., and Sung, H.-G., \enquote{Numerical Study of the Dynamic
  Characteristics of Pintle Nozzles for Variable Thrust,} \emph{JOURNAL OF
  PROPULSION AND POWER}, Vol.~31, No.~1, 2015, pp. 230--237.

\bibitem{Ko:13}
Ko, H., Lee, J.-H., Chang, H.-B., and Yoon, W.-S., \enquote{Cold Tests and the
  Dynamic Characteristics of the Pintle Type Solid Rocket Motor,} in
  \enquote{49th AIAA/ASME/SAE/ASEE Joint Propulsion Conference,} San Jose, CA,
  2013.

\bibitem{Lee:08}
Lee, J.~H., Kim, J.~K., Jang, H.~B., and Oh, J.~Y., \enquote{Experimental and
  Theoretical Investigations of Thrust Variation with Pintle Positions Using
  Cold Gas,} in \enquote{44th AIAA/ASME/SAE/ASEE Joint Propulsion Conference
  and Exhibit,} Hartford, CT, 2008.

\bibitem{Verma:11}
Verma, S.~B. and Haidn, O., \enquote{Cold Gas Testing of Thrust-Optimized
  Parabolic Nozzle in a High-Altitude Test Facility,} \emph{Journal of
  Propulsion and Power}, Vol.~27, No.~6, 2011, pp. 1238--1246.

\bibitem{Peterson:12}
Peterson, Z.~W., Eilers, S.~D., and Whitmorey, S.~A., \enquote{Closed-loop
  Thrust and Pressure Profile Throttling of a Nitrous-Oxide HTPB Hybrid Rocket
  Motor,} in \enquote{48th AIAA/ASME/SAE/ASEE Joint Propulsion Conference and
  Exhibit,} Atlanta, Georgia, 2012.

\bibitem{Ann:05}
Narendra, K.~S. and Annaswamy, A.~M., \emph{Stable adaptive systems}, Dover
  Publications, New York, 2005.

\bibitem{Lavretsky:book}
Lavretsky, E. and Wise, K.~A., \emph{Robust Adaptive Control}, Springer,
  London, 2013.

\bibitem{Yil:11}
Yildiz, Y., Annaswamy, A.~M., Yanakiev, D., and Kolmanovsky, I.,
  \enquote{Spark-Ignition-Engine Idle Speed Control: An Adaptive Control
  Approach,} \emph{IEEE TRANSACTIONS ON CONTROL SYSTEMS TECHNOLOGY}, Vol.~19,
  No.~5, 2011, pp. 990--1002.

\bibitem{Gibson:phd}
Gibson, T.~E., \emph{Closed-Loop Reference Model Adaptive Control: with
  Application to Very Flexible Aircraft}, Phd thesis, Massachusetts Institute
  of Technology, 2014.

\end{thebibliography}

\end{document}